# Emergent Anti-Ferroelectric Ordering and the Coupling of Liquid Crystalline and Polar Order

Jordan Hobbs * [1], Calum J. Gibb [2], and Richard. J. Mandle* [1,2]

[1] School of Physics and Astronomy, University of Leeds, Leeds, UK, LS2 9JT
[2] School of Chemistry, University of Leeds, Leeds, UK, LS2 9JT

*j.l.hobbs@leeds.ac.uk

## Abstract

Polar liquid crystals possess three dimensional orientational order coupled with unidirectional electric polarity, yielding fluid ferroelectrics. Such polar phases are generated by rod-like molecules with large electric dipole moments. 2,5-Disubstituted 1,3-dioxane is commonly employed as a polar motif in said systems, and here we show this to suffer from thermal instability as a consequence of equatorial-trans to axial-trans isomerism at elevated temperatures. We utilise isosteric building blocks as potential replacements for the 1,3-dioxane unit, and in doing so we obtain new examples of fluid ferroelectric systems. For binary mixtures of certain composition, we observe the emergence of a new fluid antiferroelectric phase - a finding not observed for either of the parent molecules. Our study also reveals a critical tipping point for the emergence of polar order in otherwise apolar systems. These results hint at the possibility for uncovering new highly ordered polar LC phases and delineate distinct transition mechanisms in orientational and polar ordering.

## Introduction

The nematic (N) liquid crystalline (LC) phase consists of molecules or particles which lack translational order but retain a degree of orientational ordering, with the molecules aligning along a unit vector termed the director ($\hat{\boldsymbol{n}}$) (**Fig. 1a**). The conventional N phase possesses inversion symmetry (*i.e.* the nematic director ($\hat{\boldsymbol{n}}$) is equivalent to its reciprocal, $\hat{\boldsymbol{n}}$ = -$\hat{\boldsymbol{n}}$) and so is apolar, even when formed from highly polar molecules. While subject of some interest at the turn of the 20$^{th}$ century [1], an inherently polar N phase lacking inversion symmetry (i.e. $\hat{\boldsymbol{n}}$ ≠ -$\hat{\boldsymbol{n}}$) (**Fig. 1a**) was only experimentally discovered in 2017 simultaneously and independently in two different materials: **RM734** [2,3] and **DIO** [4]. Now referred to as the ferroelectric nematic ($N_F$) phase, the emergence of a new nematic phase at equilibrium has garnered significant excitement amongst the scientific community [5–11]and has been promised to 'to remake nematic science and technology' [12]. Given the low energy cost of elastic deformation of many LC, polar variants of other 'transitional' LC phases has now been observed including ferroelectric [13–16] and antiferroelectric [16] analogues of the smectic A (SmA) phase, denoted as the $SmA_F$ and $SmA_{AF}$ phases, respectively (**Fig. 1b**), as well as more recently spontaneously chiral and tilted nematic and smectic phases [17].

LCs exhibiting polar behaviour can be considered as being constructed from multiple polar fragments, such as the 2,5-disubstituted 1,3-dioxane found in DIO (**Fig 1c**). The 2,5-disubstituted dioxane makes a modest contribution to the molecular electric dipole moment (~ 1 D at MP2/6-311+G(d,p)) and is found in materials exhibiting the $N_F$ the $N_F$, $N_X$ (sometimes labelled as $SmZ_A$ or $N_S$) [4], $N_{TBF}$ [17], $SmA_F$ [13] and the $SmC_P^H$ phase [16]. However, 2-aryl-1,3-dioxane containing LCs (such as **DIO**) are widely known to have relatively poor thermal stability yet this is seldom explained beyond noting some structural rearrangement may occur above some temperature.

Herein we present a detailed analysis of the structural rearrangement that does indeed occur for 2-aryl-1,3-dioxane containing LCs and explore some alternatives to the 1,3-dioxane unit of **DIO**. We generally observe a destabilisation of polar phases except for a single compound which exhibits a $SmA_F$ phase. Mixtures of this compound with DIO are fabricated which show an emergent $SmA_{AF}$ phase. By studying various binary mixtures exhibiting polar LC behaviour we also shed light onto the relative coupling between LC and polar order for mesogenic materials.

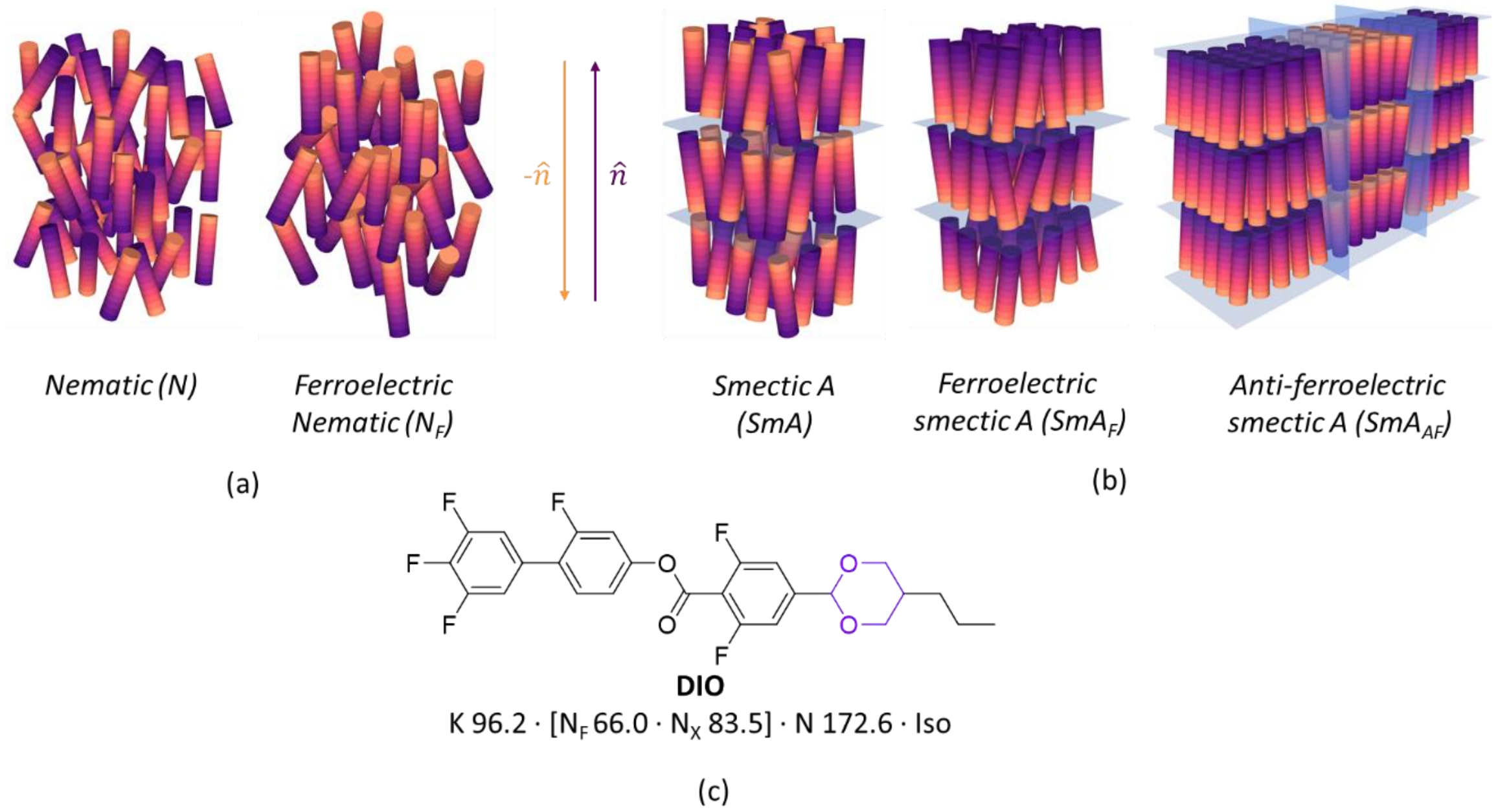

**Fig. 1. (a)** Schematic representations of the nematic (N) [left] and ferroelectric nematic ($N_F$) [right] phases, **(b)** Schematic representations of the SmA phase [left] and, an example of a lamellar phase with polar order, the $SmA_F$ phase, and **(c)** the chemical structure of the archetypal $N_F$ material DIO [4]. The 1,3-dioxane unit is highlighted in purple.

## Results and Discussion

It is known that **DIO** “degrades" when heated to temperatures above 120 °C, and this has been suggested to be a trans-cis structural isomerism concerning the 1,3-dioxane unit. [4,18] A sample of **DIO** was subjected to repeated heat/cool using DSC cycles, varying the maximum temperature (100 °C to 220 °C, 10 °C increments, **Fig. S1**). Repeated heating and cooling cycles show no decrease in the onset of the ferroelectric to anti-ferroelectric and anti-ferroelectric to nematic phase transitions ($T_{N_F\text{-}N_X}$ and $T_{N_X\text{-}N}$, respectively) until 140 °C, where subsequent cycles resulted in decreased in all the mesophase transition temperatures (**Fig. 2a**). The decrease is continuous until 190 °C where the transition temperatures plateaus at approximately 10 °C below their initial values; the plateau implies an equilibrium is attained as opposed to continuous thermal decomposition. The proposed *cis*/*trans* isomerisation of 1,3-dioxane cannot occur without breaking and remaking C-O bonds, thus we consider this unlikely at the temperatures in question. All 2,5-di-substituted 1,3-dioxanes have two principal *trans* conformations: *trans* equatorial (*eq-trans*), and *trans* axial (*ax-trans*) which can interconvert via twist-boat states and half-chair transition state conformations (**Fig. 2b**) [19]. Calculation of the intrinsic reaction coordinate pathway (IRC; at the B3LYP-GD3BJ/cc-pVTZ level of DFT) confirms the *eq-trans* to *ax-trans* assignment and supports the isomerisation proceeding by this mechanism (**Fig. 2b**). The NMR spectrum of a sample of **DIO** heated above 190 °C showed shifts in resonances consistent with partial isomerisation from the *eq-trans* form to the *ax-trans* form. Assignment of these changes in chemical shift was supported the calculated NMR shielding tensors of *ax-trans* (B3LYP-GD3BJ/cc-pVTZ level of DFT). A comparison of the peak areas of the two singlets reveals that the *ax-trans* conformation comprises only 3% of the resultant mixture after heating, consistent with the modest decrease in the values of $T_{N_F\text{-}N_X}$ and $T_{N_X\text{-}N}$ observed by DSC as well as the 7.7 kJ/mol energy difference as calculated from the DFT IRC trace.

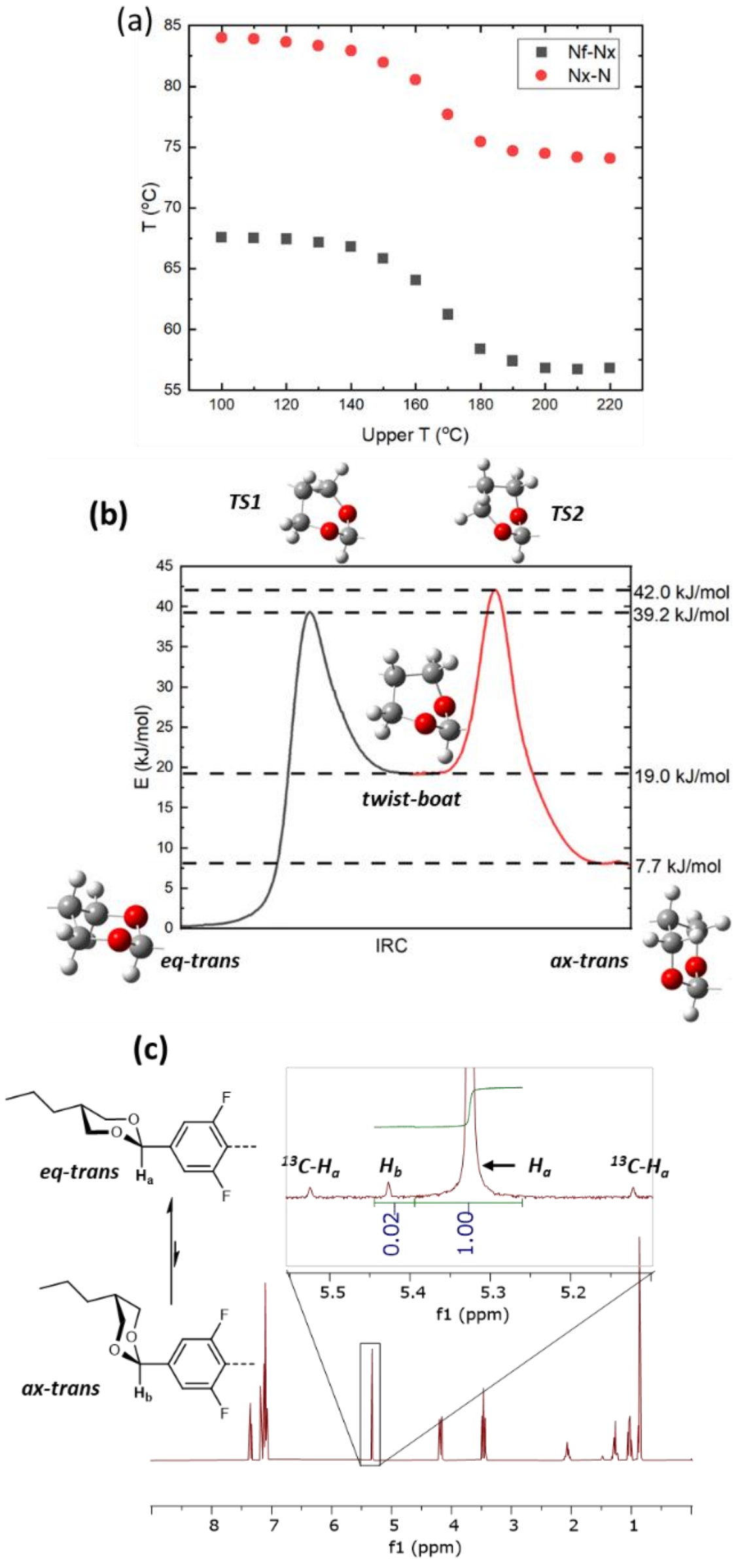

**Fig. 2. (a)** Transition temperature of **DIO** which has been heated to the temperature marked as "upper T" as determined via DSC. **(b)** IRC trace of **DIO** showing the eq-trans to ax-trans isomerisation via half-chair (TS) and twist boat forms. **(c)** NMR spectra of **DIO** which had been heated to 190 °C showing the emergence of a second peak associated with the ax-trans form of **DIO**; relevent integrated peak areas are given under each peak.

With the thermal instability of the 1,3-dioxane unit now understood, we move to study alternative ring systems with a view to enhancing the thermal stability of the parent molecule,

**DIO**. Given the number of possible end-group we could have chosen, we imposed two constraints to focus our investigation. The first is that all end-groups possess a propyl terminal chain as this has been shown to be an appropriate length to facilitate favourable head-to-tail correlations between neighbouring molecules [2,20,21] The second is that all end-groups must contribute to the overall longitudinal dipole moment of the molecule (μ) as a large value of μ is thought necessary for the formation of polar mesophases [9,22]. The terminal groups chosen, their associated transitional data, and contribution to the molecular dipole moment are shown in **Fig. 3** and are given in tabulated form in the ESI (**Table S1**)**.** A detailed description of the chemical synthesis of these materials is provided in the ESI to this article.

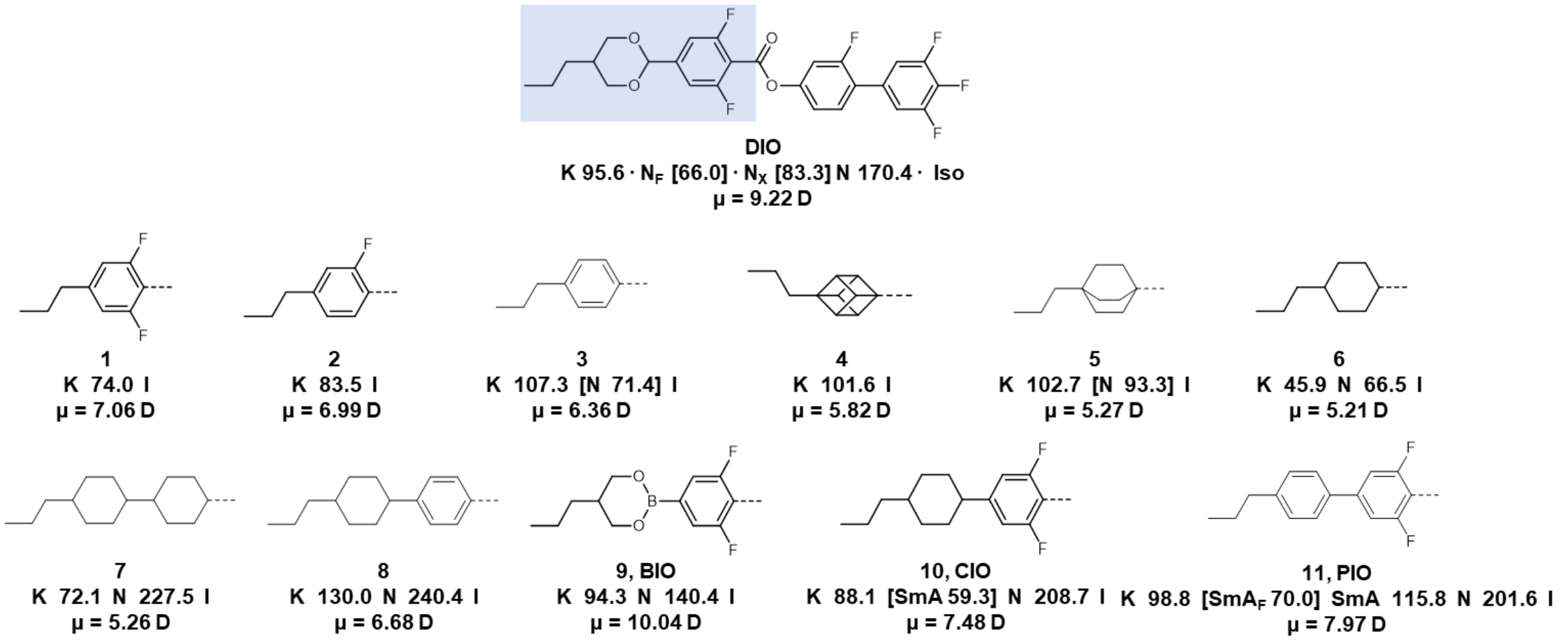

**Fig. 3.** Chemical structures of the terminal groups chosen as alternatives to the 1,3 dioxane ring seen in **DIO** for this work, their associated transitional properties, and longitudinal molecular dipole moments (μ, at the B3LYP-GD3BJ/cc-pVTZ level of DFT) of compounds **1-11**. The transition temperatures given are quoted in °C with '[ ]' indicating a monotropic phase transition. Specific compounds of interest are given short generic names (**9**: BIO, **10**: CIO, **11**: PIO).

We synthesised six materials whereby the 2-(3,5-difluorophenyl)-1,3-dioxane unit is supplanted by a 'single ring' bearing propyl terminal chain (**1-6**). These compounds variously incorporate aromatic (**1-3**), and/or fluorinated (**1**, **2**), saturated (**4-6**) and/or bi/polycyclic (**4**, **5**) ring systems. Compound **1** can be considered as DIO minus the 1,3-dioxane ring, however **1** is non mesogenic as is the closely related compound **2**. The closely related compound **3**, differing from **1** and **2** in terms of the number of fluorine atoms on the benzoate unit, exhibits a monotropic nematic phase. Replacement of the 1,4-benzene ring of **3** with other isosteric groups gave mixed results: the 1,4-cubane (**4**) is non mesogenic; the 1,4-bicyclo[2.2.2]octane (**5**) and the *trans* 1,4-cyclohexane (**6**) are both nematic. Compounds **3**, **5** and **6** exhibit nematic phases with significantly reduced clearing points relative to the parent **DIO**, and we conjecture that this is related to the decreased molecular length.

We therefore prepared materials with similar aspect ratios to **DIO** by incorporating an additional ring unit. Compound **7** is related to **6** but bears and additional *trans* 1,4-cyclohexane unit and exhibits a considerable increase in $T_{N\text{-}I}$ (over 160 °C) with only a modest increase in melting point. Similarly, compound **8** is analogous to **3** but with the addition of a *trans* 1,4-cyclohexane (or alternatively, analogous to **6** but with an additional 1,4-benzene); dramatic increases in clearing point are observed, as well as significantly higher melting point. For both **7** and **8,** the molecular electric dipole moments are comparable with the parent 3-ring systems (**3** or **6**) and the mesophases observed are non-polar. Next, we sought to increase the polarity of the materials (and the molecular electric dipole moment) by use of heterocycles and/or fluorination. Compound **9** (aka **BIO**) bears a 1,3,2-dioxaborinane ring as well as a 3,5-difluorobenzene; this combination yields a larger molecular dipole moment than **DIO**. However, **BIO** both unstable with respect to air and water as well as only exhibiting solely conventional nematic behaviour.

The 1,3-dioxane of DIO is replaced with cyclohexane in compound **10** (aka **CIO**). While this leads to a somewhat lower molecular electric dipole moment (7.48 D versus 9.22 D for **DIO**, both at the B3LYP-GD3BJ/cc-pVTZ level of DFT), **CIO** has a significantly higher clearing point (and a slightly lower melting point) than **DIO**. We find that **CIO** exhibits a monotropic conventional SmA phase below an enantiotropic N phase which was characterised by the observation of focal conic fans (**Fig. S4**) by POM. We again however see no evidence of any polar mesophase behaviour for **CIO** in either POM observations or Ps measurements despite possessing a similar molecular shape and a comparable molecular dipole moment. Work by Madhusudana has previously suggested that a longitudinal surface charge density wave along the length of the molecule is important in stabilizing parallel molecular orientations. [23] The limited anisotropic polarizability of the cyclohexane moiety in **CIO** may therefore be detrimental to the formation of polar mesophases. The electrostatic potential energy surface was calculated using B3LYP-GD3BJ/cc-pVTZ level of DFT. By radially averaging and plotting the electrostatic potential as a function of the z-axis (described in more detail in the methods) we can see that electronically speaking, the cyclohexane ring acts essentially as an extension of the terminal chain reducing the prominence of the longitudinal surface charge density wave (**Fig. 4a, b**). Replacing the cyclohexane moiety with a phenyl ring (compound **11** aka **PIO**) introduces additional anisotropic polarizability, returns the oscillatory charge density wave found for **DIO** (**Fig. 4c**, **Fig S5**) while the increased conjugation leads to a somewhat higher molecular electric dipole moment, and so we considered this to be a promising candidate for exhibiting polar LC phases. We find **PIO** to have a slightly higher $T_{NI}$ and $T_m$ than **DIO**. Significant stabilisation of smectic ordering is found in **PIO**, with a transition from the N to a SmA phase at 115.8 °C, an increase of around 55 °C compared **CIO**. The assignment of the SmA phase was confirmed by POM via the observation of a focal conic fan texture (**Fig. 4d**) and subsequent Bragg scattering in the SAXS pattern (**Fig. 4e**).

Given its similarity to **DIO**, the lack of polar phase behaviour in **PIO** was surprising. DSC measurements and POM cooling at rate of around 10 °C/min resulted in crystallisation around 80 °C giving only a small window of supercooling to observe any further phase transitions. We hypothesised that any transitions to a polar phase would be at least below 80 °C (the transition to the $N_X$ phase in **DIO**). We therefore studied **PIO** by POM and measured its response to spontaneous polarisation ($P_S$) while rapidly cooling the sample (>100 °C/min). At approximately 70 °C, a transition to polar phase was observed which we assign to be a $SmA_F$ phase by the appearance of a blocky mosaic texture (**Fig. 4d**) and a single peak in the current response (**Fig. 4f**). These observations are consistent with previous observations of the $SmA_F$ [13,14,16]Ideally, this phase assignment would be verified via X-ray scattering and detailed $P_S$ measurements however, the stringent cooling requirements prohibited our analysis and so we envisioned the formulation of binary mixtures with **DIO** would suppress the melting points sufficiently, allowing for this phase assignment to be verified.

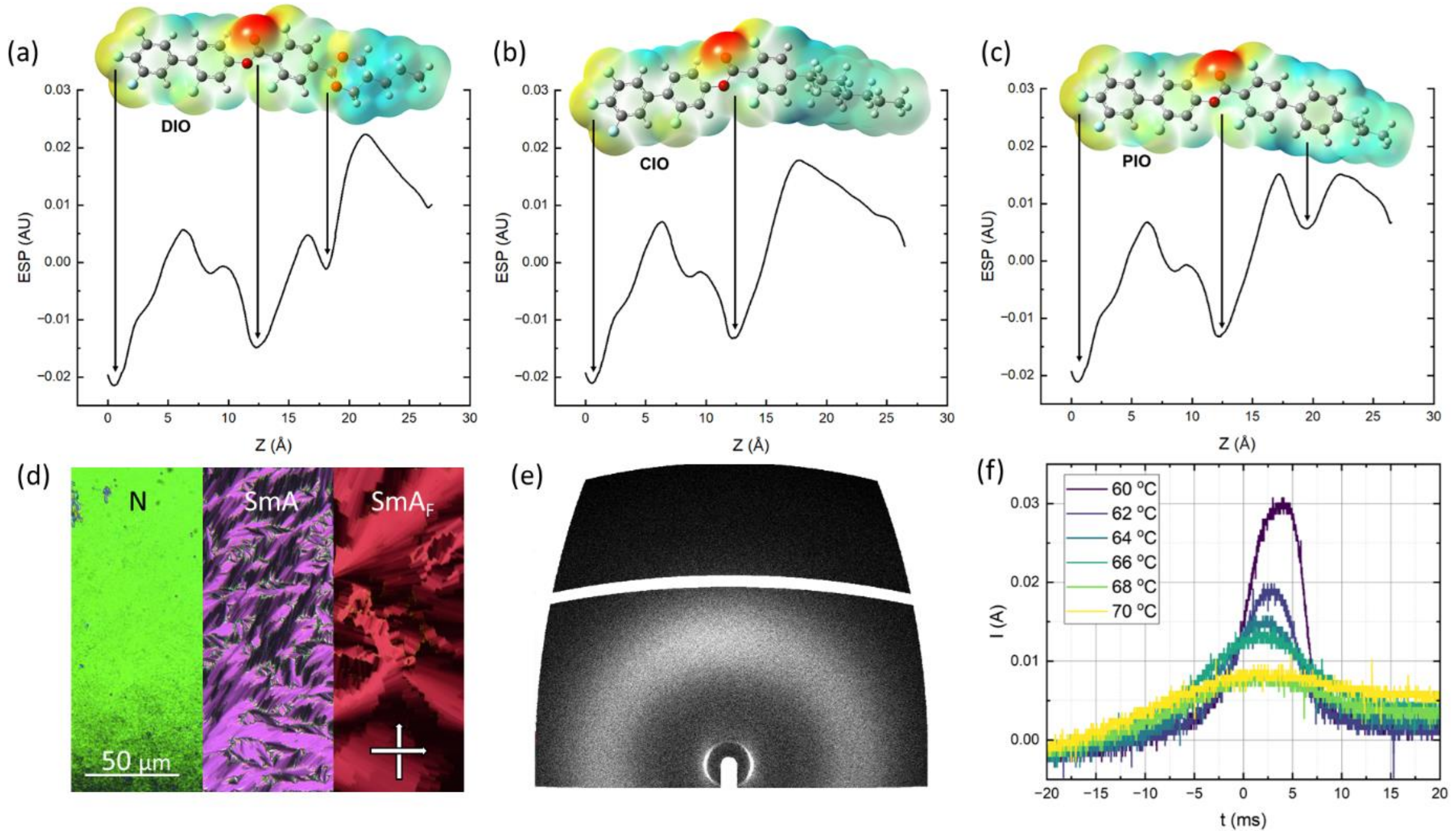

**Fig. 4.** Electrostatic potential energy along the z-axis of the molecule at an electron density iso-surface of 0.0004 calculated at the B3LYP-GD3BJ/cc-pVTZ level of DFT for **(a) DIO**, **(b) CIO**, and **(c) PIO**. **(d)** POM micrographs depicting the phase sequence of **PIO**. Images were taken within 8µm cells with no alignment layer**, (e)** 2-D X-ray scattering pattern of **PIO** in the SmA phase and **(f)** current response of **PIO.** Given the propensity for crystallisation of the sample, rapid cooling of the sample (>100 °C) is necessary to study the phase. The single polarization peak indicates the ferroelectric nature of **PIO** at ~70 °C although the magnitude of the peak is unreliable due to partial sample crystallisation.

Mixing **PIO** and **DIO** (**Fig. 5(a)**) results in a modest reduction in $T_m$ (~20 °C reduction from **DIO**) leading to several significant observations. In high concentrations of **DIO,** phase behaviour matching that of the dominant molecule is observed which, when combined with the reduction in the values of $T_m$, leads to mixtures showing enantiotropic $N_F$ phases which could be supercooled to room temperature in as little as 10 mol% **PIO**. The $N_F$ phases were characterised by the observation of banded textures using POM (**Fig. 5(b)**), seemingly characteristic of the $N_F$ phase, with the assignment further affirmed by the measurement of spontaneous polarization (**Fig. 5 (c)**). For mixtures comprising of approximately equal amounts of **PIO** and **DIO** (30-60 mol% **DIO**), a phase with a similar blocky fan POM texture (**Fig. 5(d)**) is observed, similar to that seen for pure **PIO**. WAXS and SAXS measurements confirm liquid like order with a lamellar structure, with the layer normal parallel to the director (**Fig. 5(f)**). $P_S$ measurements confirm that the phase is ferroelectric from the single peak in the current response (**Fig. 5(f)**). Extrapolation of these transitions shows a linear dependence with the $SmA_F$ phase transition assigned for pure **PIO**.

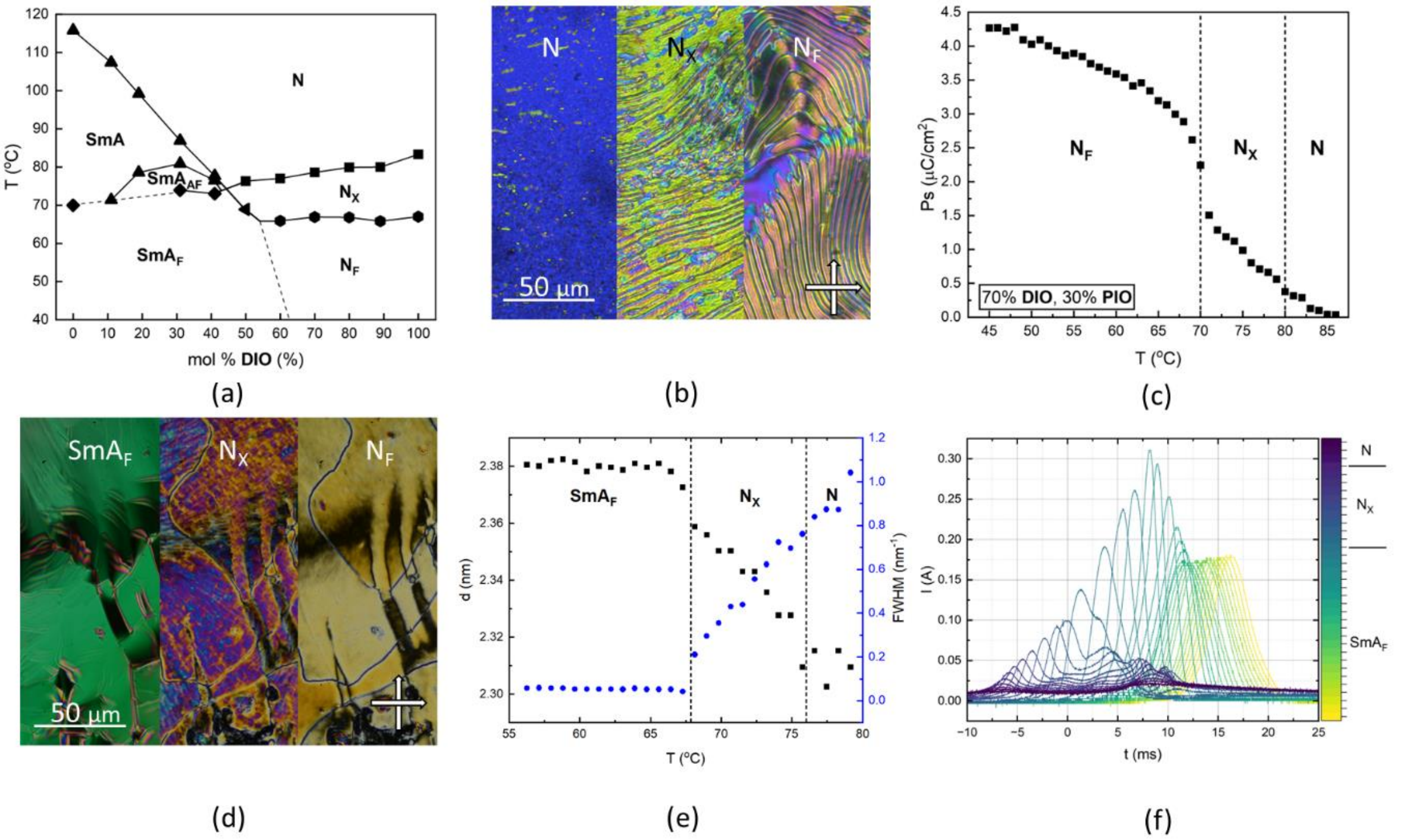

**Fig. 5. (a)** Phase diagram for mixtures of **PIO** with **DIO**, **(b)** POM micrographs depicting the phase sequence the mixture comprised of 70% **DIO** and 30% **PIO**. Images were taken within 8 μm cells with no alignment layer, **(c)** variation of $P_S$ as a function of temperature measured for the mixture comprised of 70% **DIO** and 30% **PIO.**, **(d)** POM micrographs depicting the blocky texture of the SmAF phase observed for a mixture comprising of 50% **DIO** and 50% **PIO.** Images were taken within 8 μm cells with no alignment layer**. (e)** Position and FWHM of the X-ray scattering peak associated with long molecular axis corelations. **(f)** Spontaneous polarization current response of mixtures containing 50% **DIO** and 50% **PIO**.

Mixtures fabricated from **PIO** and **DIO**, allow for significant supercooling of the SmA$_F$ phase below their melting points (circa >15 °C below $T_m$ by POM). For mixtures comprised of between 10 – 40 mol% **DIO**, conventional N and SmA phases are observed, with a phase transition to a new smectic phase not observed for either **PIO** or **DIO**. Texturally the phase presents with a similar texture to the SmA phase by POM (as described previously) however, a small step-like transition with a small enthalpy change is observed in the DSC thermograms (**Fig. 6(a)**). SAXS experiments indicates that the phase has a similar lamellar structure to the preceding SmA phase, with no change in the layer spacing observed at the phase transition (**Fig. 6(b)**). Measurements of the current response show a transition from apolar ordering in the SmA phase to a double peak upon cooling into the lower temperature phase, indicating anti-ferroelectric ordering within the lamellar structure (**Fig. 6(c)**). Further cooling this phase sees the recombining of the double peak into a single peak (ferroelectric-like ordering) with the blocky focal conic fan texture of the SmA$_F$ phase observed by POM. A longitudinal anti-ferroelectric SmA (SmA$_{AF}$) phase has been demonstrate previously in highly polar LCs [16] though the emergence of anti-ferroelectric ordering here is certainly unexpected. No evidence for phase separations can be seen in either POM or SAXS measurements. Ferroelectric order can be induced in N materials above the transition to polar order [24] using electric fields which could explain the anti-ferroelectric response (peak on increasing voltage associated with induced order while the one on reducing voltage associated with the return to apolar order) however, the distinct thermal transition from the DSC suggests that this is also not the case. At a concentration of 40% **DIO**, the transitions from N to SmA to SmA$_{AF}$ all occur in a small temperature window and as such additional pre-transitional anti-ferroelectric ordering can be seen in the N phase (as it transitions to the SmA and subsequent $SmA_{AF}$ phase) in the form of chevron defects (**Fig. 6(d**), which have been observed to occur in the $N_X$ phase [25] which suggests that the appearance of anti-ferroelectric order may also be unrelated to the lamellar order of the SmA phase. Typically speaking, in apolar liquid lamellar crystals such as 8CB the replacement of a 1,4-disubstituted benzene ring with a 2,5-disubstituted 1,3-dioxane supresses the formation of smectic layers, giving only a nematic phase [26]; the same behaviour would appear to be at play here [26]. While we are not certain as to the molecular origins of $SmA_{AF}$ phase in this case, we suggest that **DIO** and **PIO** possibly have some molecular interaction which promotes the formation of the splay domains within this specific concentration region on top of the molecule's inherent tendency to form polar LC phases.

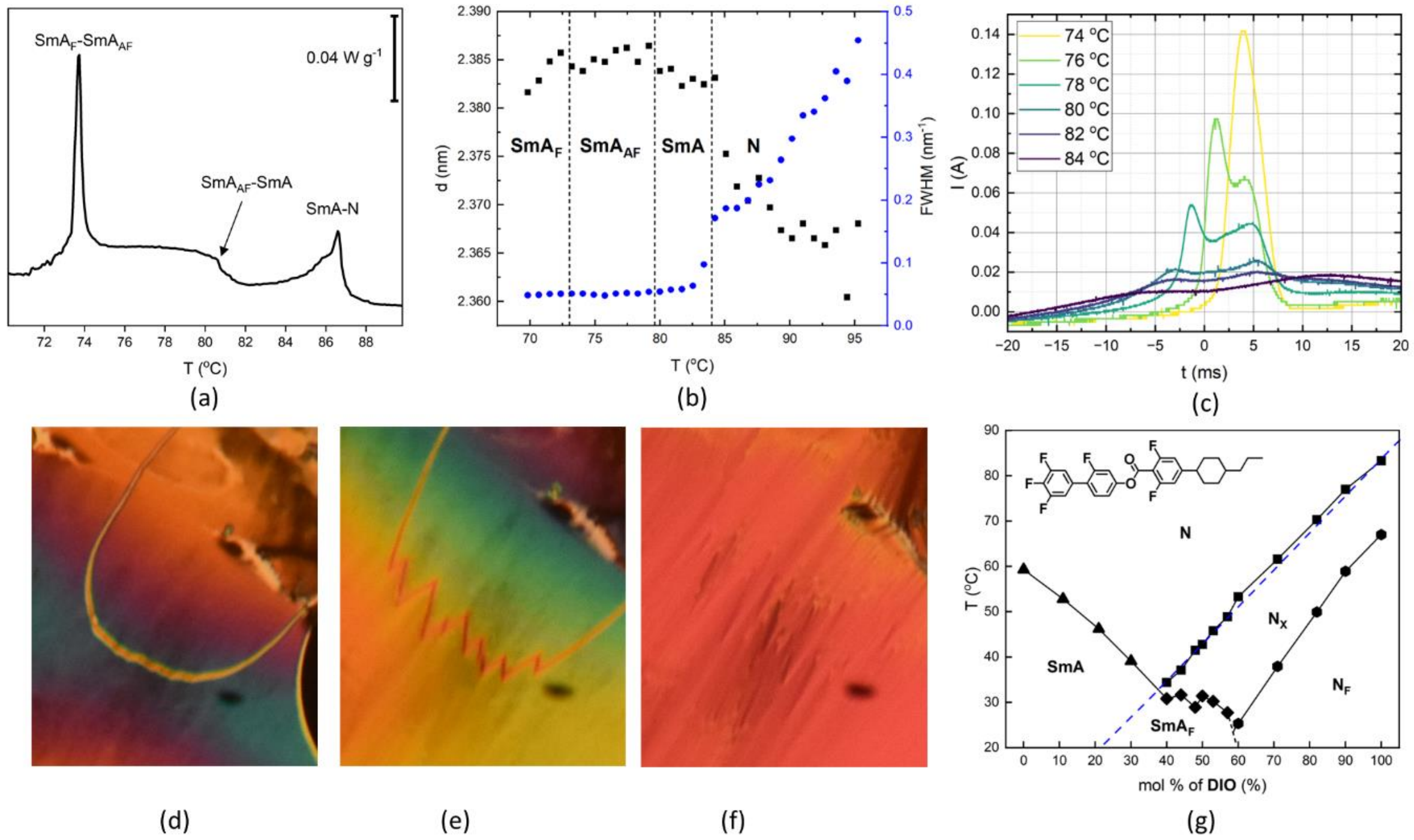

**Fig. 6 (a)** DSC thermogram for 30% **DIO** and 70% **PIO** showing the thermal transition into the $SmA_{AF}$ phase. **(b)** Position and FWHM of the X-ray scattering peak associated with long molecular axis corelations for 30% **DIO** and 70% **PIO**. **(c)** Current response for the $SmA_{AF}$ phase showing the two peaks. **(d), (e)** and **(f)** POM micrographs of a mixture of 40% **DIO** and 60% **PIO** showing how a chevron defect evolves between the narrow ~2 °C window where the sample transitions from N-SmA-$SmA_{AF}$. **(g)** Phase diagram between **DIO** and **CIO** with the transition from apolar-polar order marked in blue.

For binary mixtures of **DIO** with either **PIO** or **CIO** we find the onset temperature of polar order to vary linearly (to a first approximation). A linear line can be drawn from the N-$N_X$ transition in **DIO** to the SmA-$SmA_F$ transition in **PIO,** passing through the transition to $N_X$ and $SmA_F$ phases across the mixture diagram (**Fig. 6(e)**). To the authors knowledge, this holds true for all published binary mixtures involving two longitudinally polar LCs [16,27–29]. This implies that some temperature exists for a material at which polar ordering of the molecules will be preferable [23], and below this temperature, polar LC phases may be formed regardless of the underlying preferred LC phase. Many factors would influence this temperature, not least combinations of electrostatic and steric contributions.

We suggest that in binary mixtures between a LC that shows polar phases and a LC that does not (at least not before crystallisation though we do accept that there may be some molecules fundamentally incapable of forming longitudinally polar LC phases), this linear dependence is still true but will be between the apolar-polar transition in the polar LC and a virtual apolar-polar transition in the non-polar LC. For materials, such as 5CB or E7, this virtual transition will be very low and so in binary mixtures the transition to polar order quickly drops out of the phase diagram [30,31]. To test this hypothesis, we elected to make mixtures of **DIO** with **CIO** (**Fig. 6(f)**) and **BIO** (**Fig. S7**), as the parent molecules did not show any evidence of polar phases. Both materials demonstrated linear dependence on molar concentration for the transition to polar order, enabling a "virtual" transition to a polar phase to be extrapolated which for **BIO** is -1.6 °C and for **CIO** is 1.5 °C. **BIO** shows only a N phase and so the phase diagram with **DIO** shows only N type phases. The SmA phase exhibited by **CIO** makes its phase diagram with **DIO** is more complex. In high concentrations of either **CIO** or **DIO**, the mixtures behave like the parent molecules, with the expected reduction in transition temperatures as the concentration of the counter molecule is increased. Approaching the eutectic point (40-60% **DIO**) where the transition to a SmA phase crosses the transition to polar order results in a region where a $SmA_F$ phase is formed.

From studying the binary mixtures presented here and elsewhere [16] the empirical linear dependence of the apolar-polar transition for binary mixtures seems to hold regardless of whether it is a polar N phase or a polar Sm phase being formed, i.e. there is no apparent interdependence of longitudinal polar and LC order beyond orientational order. This suggests, generally speaking, that the mechanism that drives the LCs to exhibit longitudinal polar order is at least partially separate to the mechanisms that drive a LC to form either N, SmA or SmC phase. We suggest that "double" transitions (ones where there is a change in LC order AND polar order i.e. N-$SmA_F$) are improbable for this reason, although there is no *a priori* reason this should not occur. We do accept that a transition in LC order could occur close enough to a transition in polar order to appear as if it is one transition [32]] though we suggest that there are separate underlying mechanisms driving each element of this double transition as separate entities. The only requirement of orientational order suggests the possibility of generating polar equivalents of most (if not all) LC phases exhibited by calamitic like molecules (such as SmB or SmG for example) and indeed recently molecules showing polar SmC phases have been demonstrated [16]]. Moreover, the interplay between escaping polarity and positional order could be used to generate as yet unseen phase types. This is a significantly different situation from other polar LC phases such as those found in SmC* [33] or bent-core polar LCs [34] where the polarity of the system is fundamentally coupled to the LC phase.

## Conclusion

Several highly polar liquid crystalline materials were synthesised as a means to escape the thermal isomerisation we observe for the 2,5-disubstituted 1,3-dioxane motif in the widely used $N_F$ host material **DIO**. Although we find only ~3% of the material isomerises, this is sufficient to significantly depress phase transitions and presents a significant barrier to working with this material. We developed the material "**PIO**", finding it to exhibit a $SmA_F$ phase. Remarkably, binary mixtures of **PIO** and **DIO** showed an injected antiferroelectric phase ($SmA_{AF}$). Using phase diagrams of **DIO** with **PIO**, **CIO** and **BIO**, as well as phase diagrams from literature, we show that calamitic LC materials have some temperature at which they will form longitudinal polar LC phases. This apolar-polar transition temperature is independent of the underlying LC phase present as the only requirement for polar order is orientational order which separates the transition into two types: those where LC order changes (e.g., N-SmA or $N_F$-$SmA_F$) and those where the polar order changes (e.g., N-$N_F$ or SmA-$SmA_F$).

## Data availability

All data needed to evaluate the conclusions in the paper are present in the paper and/or the Supplementary Materials. The raw data and associated metadata associated with this paper are openly available from the University of Leeds Data Repository at https://doi.org/10.5518/1510.

## Acknowledgements

RJM acknowledges funding from UKRI via a Future Leaders Fellowship, grant no. MR/W006391/1, and funding from the University of Leeds via a University Academic Fellowship. The SAXS/WAXS system used in this work was funded by EPSRC *via* grant number EP/X0348011. R.J.M. gratefully acknowledge support from Merck KGaA.

## Author contributions

CJG and R.J.M performed chemical synthesis; J.L.H and C.J.G. performed microscopy, J.L.H performed X-ray scattering experiments, applied field studies, mixture formulation studies and DSC analysis; J.L.H and R.J.M. performed and evaluated electronic structure calculations; R.J.M secured funding. The manuscript was written, reviewed and edited with contributions from all authors.

# Emergent Anti-Ferroelectric Ordering and the Coupling of Liquid Crystalline and Polar Order

## Supplemental Information

Jordan Hobbs * [1], Calum J. Gibb [2], and Richard. J. Mandle* [1,2]

[1] School of Physics and Astronomy, University of Leeds, Leeds, UK, LS2 9JT
[2] School of Chemistry, University of Leeds, Leeds, UK, LS2 9JT

*j.l.hobbs@leeds.ac.uk

**Contents:**

### 1. Experimental Methods

### 1.1. Chemical Synthesis

Chemicals were purchased from commercial suppliers (Fluorochem, Merck, Ambeed, Specs) and used as received. Solvents were purchased from Merck and used without further purification. Reactions were performed in standard laboratory glassware at ambient temperature and atmosphere and were monitored by TLC with an appropriate eluent and visualised with 254 nm light. Chromatographic purification was performed using a Combiflash NextGen 300+ System (Teledyne Isco) with a silica gel stationary phase and a hexane/ethyl acetate gradient as the mobile phase, with detection made in the 200-800 nm range. Chromatographed materials were subjected to re-crystallisation from an appropriate solvent system.

### 1.2 Chemical Characterisation Methods

The structures of intermediates and final products were determined using $^{1}H$, $^{13}C\{^{1}H\}$, and $^{19}F$ NMR spectroscopy. NMR was performed using a Bruker Avance III HDNMR spectrometer operating at 400 MHz, 100.5 MHz or 376.4 MHz ($^{1}H$, $^{13}C\{^{1}H\}$ and $^{19}F$, respectively). Unless otherwise stated, spectra were acquired as solutions in deuterated chloroform, coupling constants are quoted in Hz, and chemical shifts are quoted in ppm.

### 1.3 Mesophase Characterisation

Transition temperatures and measurement of associated latent heats were measured by differential scanning calorimetry (DSC) using a TA instruments Q2000 heat flux calorimeter with a liquid nitrogen cooling system for temperature control. Between 3-8 mg of sample was placed into T-zero aluminium DSC pans and then sealed. Samples were measured under a nitrogen atmosphere with 10 °C min$^{-1}$ heating and cooling rates. The transition temperatures and enthalpy values reported are averages obtained for duplicate runs. In general LC phase transition temperatures are measured on cooling from the onset of the transition while melt temperatures were measured on heating to avoid crystallization loops that can occur on cooling. Phase identification by polarised optical microscopy (POM) was performed using a Leica DM 2700 P polarised optical microscope equipped with a Linkam TMS 92 heating stage. Samples were studied sandwiched between two untreated glass coverslips.

## 1.4 DFT Calculations

Electronic structure calculations were performed using Gaussian G16 revision C.02 [35] and with a B3LYP-GD3BJ/cc-pVTZ [36–38]basis set. For each input structure we used the ETKDGv3 rules-based method [39]to generate the lowest energy conformer for a given chemical structure. The conformer then underwent geometry optimisation followed by a frequency calculation to confirm the geometry to be at a minimum.

To obtain the optimised geometry of the various conformers of **DIO** studied here we took the lowest energy conformer which was equatorial trans **DIO** (this is “standard” **DIO**) and altered the structure of the dioxane ring to obtain both axial-trans **DIO** and the stable intermediate twist-boat state between the trans axial and equatorial forms. Both were verified as a minimum from frequency calculations. Transition states were found from estimating the “likely” medium position between the initial and final states and then optimising. The force constants were calculated at the beginning of the optimisation step and then not recalculated. The molecule was verified as a transition state from frequency calculations where only a single imaginary frequency was observed. It was further confirmed that this was the correct transition state by visualizing the vibration associated with the imaginary frequency and then calculating the intrinsic reaction coordinate (IRC) pathway and visually verifying that the start and end points correspond to the correct structures. For the IRC calculations the force constants were computed for the initial point only. The local quadratic approximation was used for the predictor step and 150 points were calculated along the reaction path in each direction.

"Electrostatic potential (ESP) surfaces were calculated by using the *formchk* and *cubegen* utilities. Both the electron density and ESP cube files were calculated using “fine” data resolution. The ESP surface was displayed at an electron density iso-surface of 0.0004. The 3D data was reduced into 1D by taking each plane across the z-axis given in the cube file, finding the isoline across each plane where the electron density equals 0.0004, and then averaging all ESP values that lie along that loop. We justify this averaging step through the assumption that the molecule will experience nearly free rotation around it’s z-axis."

## 1.5 X-ray Scattering

X-ray scattering measurements, both small angle (SAXS) and wide angle (WAXS) where recorded using an Anton Paar SAXSpoint 5.0 beamline machine. This was equipped with a primux 100 Cu X-ray source with a 2D EIGER2 R detector. The X-rays had a wavelength of 0.154 nm. Samples were filled into thin-walled quartz capillaries 1 mm thick. Temperature was controlled using an Anton Paar heated sampler with a range of -10 ℃ to 110 ℃ and the samples held in a chamber with an atmospheric pressure of <1 mBar. Samples were held at 110 ℃ to allow for temperature equilibration across the sample and then slowly cooled while stopping to record the scattering patterns.

No external alignment technique was used and so these measurements should be considered as “powder” samples. It should be noted that some spontaneous alignment of the LCs within the capillaries did occur leading to the classic “lobe” pattern seen in the 2D patterns. 1D patterns were obtained by radially integrating the 2D SAXS patterns. Peak position and FWHM was recorded and then converted into d spacing following Bragg’s law.

### 1.6 Measurement of Spontaneous Polarization ($P_S$)

Spontaneous polarisation measurements are undertaken using the current reversal technique [40,41]. Triangular waveform AC voltages are applied to the sample cells with an Agilent 33220A signal generator (Keysight Technologies), and the resulting current outflow is passed through a current-to-voltage amplifier and recorded on a RIGOL DHO4204 high-resolution oscilloscope (Telonic Instruments Ltd, UK). Heating and cooling of the samples during these measurements is achieved with an Instec HCS402 hot stage controlled to 10 mK stability by an Instec mK1000 temperature controller. The LC samples are held in 4µm thick cells with no alignment layer, supplied by Instec. The measurements consist of cooling the sample at a rate of 1 Kmin$^{-1}$ and applying a set voltage at a frequency of 10 Hz. The voltage was set such that it would saturate the measured $P_S$ and was determined before final data collection.

There are three contributions to the measured current trace: accumulation of charge in the cell ($I_c$), ion flow ($I_i$), and the current flow due to polarisation reversal ($I_p$). To obtain a $P_S$ value, we extract the latter, which manifests as one or multiple peaks in the current flow, and integrate as:

$$P_S = \int \frac{I_p}{2A} dt \qquad \textbf{(2)}$$

where A is the active electrode area of the sample cell. For the N, $N_X$ and, to a lesser extent, the $N_F$ phase, significant amounts of ion flow is present. For materials and mixtures that showed a paraelectric N phase followed by the anti-ferroelectric $N_X$ phases, the N phase always showed some pre-transitional polarisation as well as the significant ion flow mentioned previously. Since the following phase was anti-ferroelectric, this pre-transitional polarisation was anti-ferroelectric in character and so was decoupled from the ion flow in the same way as the $N_X$ phase and as such the $P_S$ of the N and $N_X$ phases was obtained by integrating the peak least affected by ion flow and then doubled to get the total area under both peaks [25]. For the $SmA_F$ phase found in these materials generally we observed low charge accumulation and ion flow allowing for the baseline to be easily defined and the integrated area of the peak or peaks to be obtained accurately.

## 2. Supplemental Figures

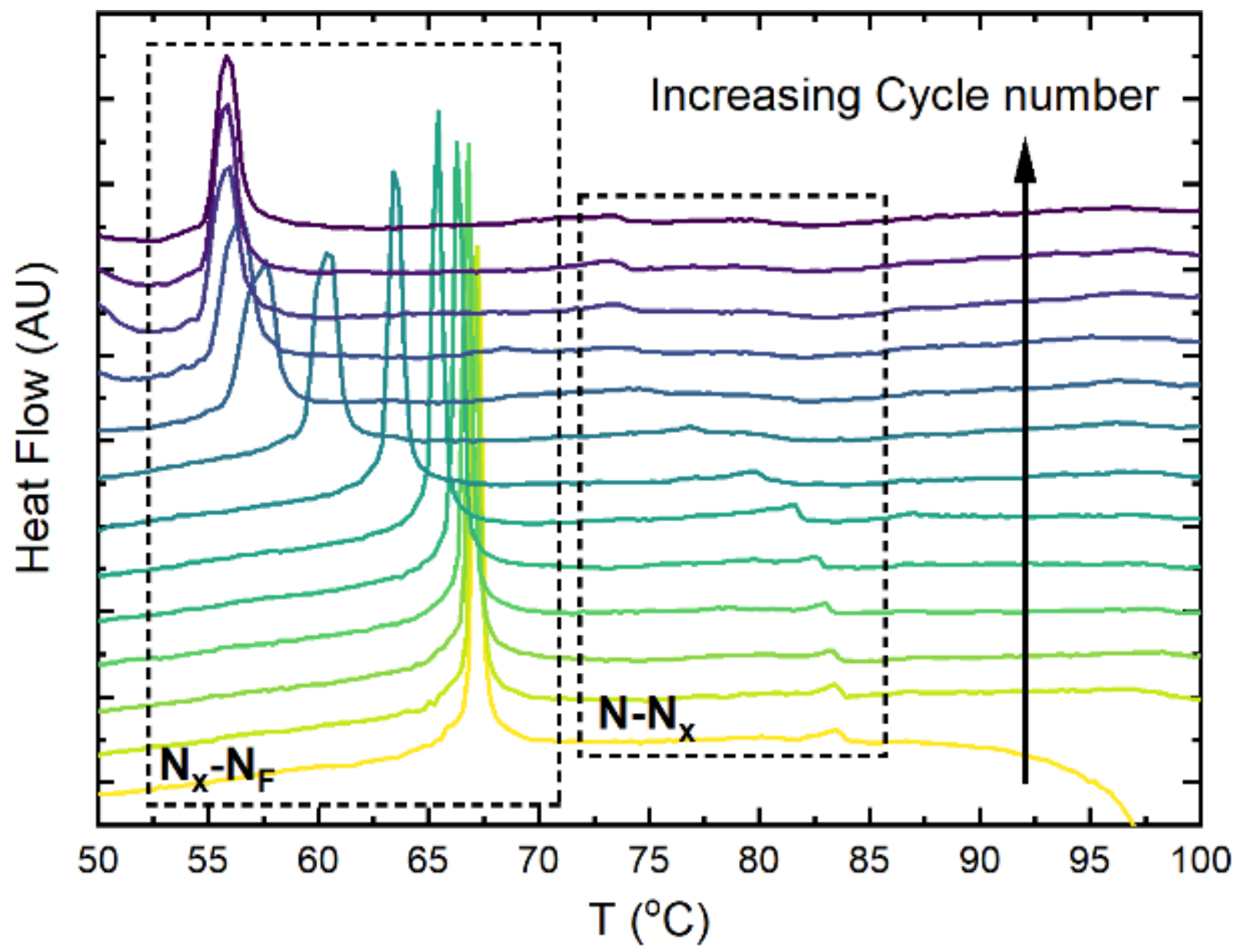

**Fig. S1** DSC thermograms of **DIO** where for each successive run the upper temperature of the run was increased by 10 °C. The direction of the arrow and colour gradient of the DSC cycles (yellow to purple) show the direction of increasing cycle number. The cycles are measured from an upper T of 100 °C to 220 °C showing the step like behaviour of the transitions due to *equatorial trans* to *axial trans* isomerisation.

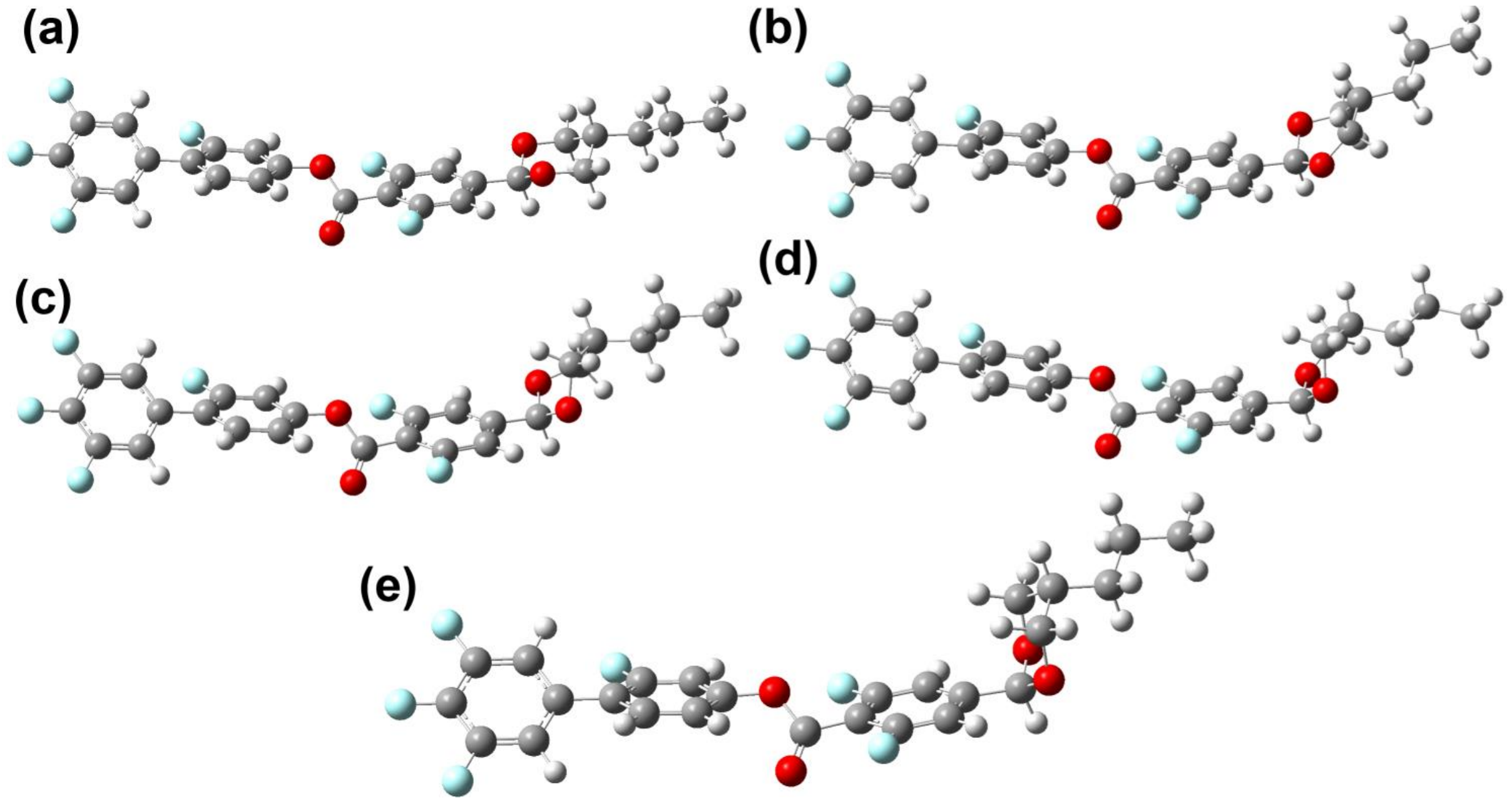

**Fig. S2** DFT optimised structures of **DIO** in the a) *equatorial trans* (eq-trans), b) 1st half-chair transition state (TS1), c) twist boat (TB), d) 2nd half-chair transition state (TS2) and e) *axial trans* (ax-trans) forms. All calculations conducted at B3LYP-GD3BJ/aug-cc-pVTZ level of DFT.

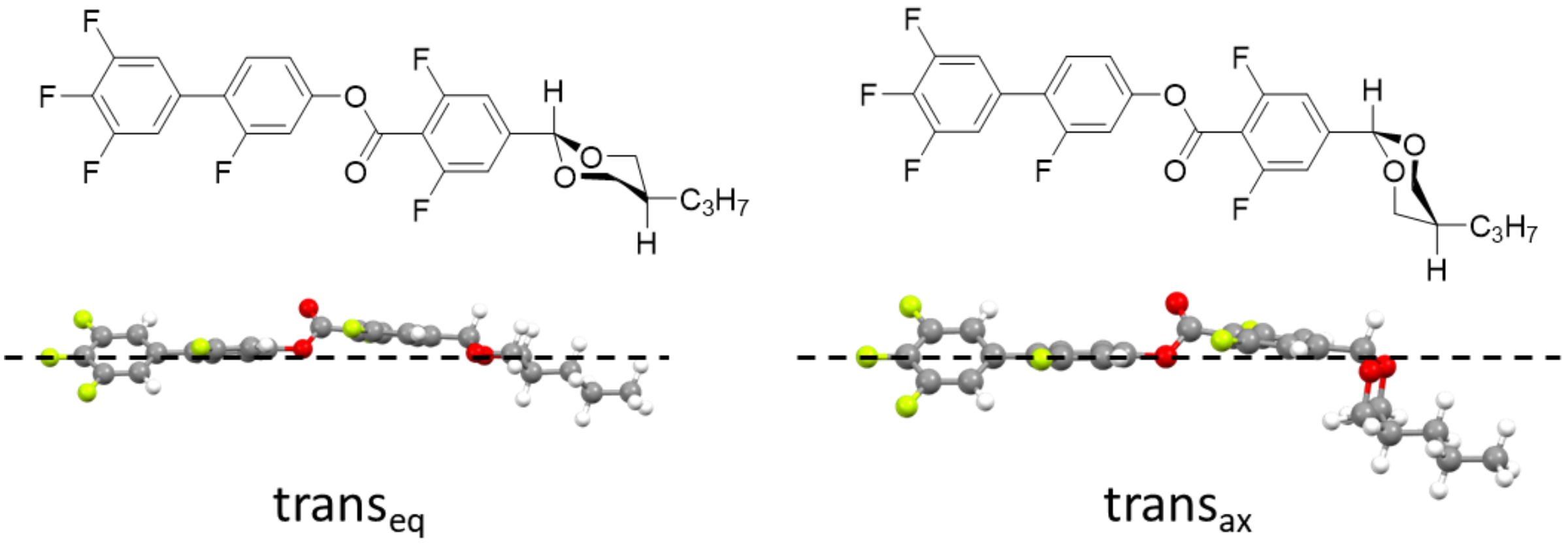

**Fig. S3** Chemical structures and DFT optimised ground states of equatorial and axial trans **DIO;** the significant deviation from linearity leads to the observed reduction in transition temperatures.

**Table S1. Tabulated transition temperatures and associated enthalpies of transition.**

| | K | | $SmA_F$-SmA | | SmA-N | | N-Iso | |
|---|---|---|---|---|---|---|---|---|
| | T (ºC) | ΔH (kJ mol$^{-1}$) | T (ºC) | ΔH (kJ mol$^{-1}$) | T (ºC) | ΔH (kJ mol$^{-1}$) | T (ºC) | ΔH (kJ mol$^{-1}$) |
| **1** | 74.0 | 30.0 | - | - | - | - | - | - |
| **2** | 85.3 | 25.4 | - | - | - | - | - | - |
| **3** | 107.3 | 30.0 | - | - | - | - | [71.4] | 0.3 |
| **4** | 101.6 | 34.2 | - | - | - | - | - | - |
| **5** | 102.7 | 23.4 | - | - | - | - | [93.3] | 0.3 |
| **6** | 45.9 | 14.7 | - | - | - | - | 66.5 | 0.3 |
| **7** | 72.1 | 17.2 | - | - | - | - | 227.5 | 0.6 |
| **8** | 130.0 | 26.7 | - | - | - | - | 240.4 | 0.6 |
| **9** | 94.3 | 22.2 | - | - | - | - | 140.4 | 0.4 |
| **10** | 88.1 | 25.1 | - | - | [59.3] | 0.05 | 208.7 | 0.6 |
| **11** | 98.8 | 23.7 | 70.0† | N/A | 115.8 | 0.2 | 201.6 | 0.5 |

† determined by POM studied, no enthalpy given; '[ ]' indicates a monotropic phase transition.

**Table S2. Molecular electric dipole moment and angle and molecular geometric parameters at the B3LYP-GD3BJ/cc-pVTZ level of DFT**

| | Dipole (D) | Dipole Angle (º) | Length (Å) | Width (Å) | Aspect Ratio |
|---|---|---|---|---|---|
| **1** | 7.06 | 12.8 | 19.03 | 5.09 | 3.74 |
| **2** | 6.99 | 5.9 | 18.91 | 4.71 | 4.02 |
| **3** | 6.36 | 13.2 | 18.88 | 4.89 | 3.86 |
| **4** | 5.82 | 12.9 | 18.64 | 5.28 | 3.53 |
| **5** | 5.27 | 13.4 | 18.57 | 5.56 | 3.34 |
| **6** | 5.21 | 12.8 | 18.88 | 4.91 | 5.21 |
| **7** | 5.26 | 16.1 | 23.92 | 5.39 | 5.26 |
| **8** | 6.68 | 9.6 | 23.03 | 6.06 | 6.68 |
| **9** | 10.04 | 9.5 | 23.61 | 5.49 | 10.04 |
| **10** | 7.48 | 10.5 | 23.12 | 6.08 | 7.48 |
| **11** | 7.97 | 10.2 | 23.09 | 5.51 | 7.97 |

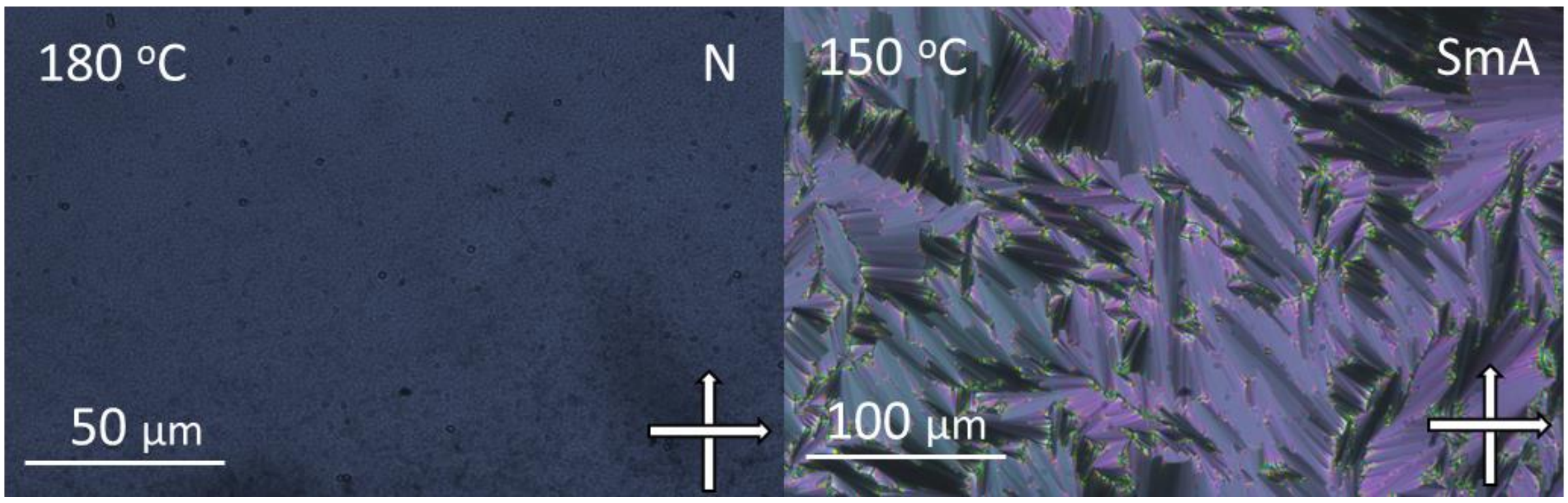

**Fig. S4** POM micrographs of a planar texture and focal conic fan texture of the N and SmA phases, respectively, observed for **10** observed within a 5 µm cell with no anchoring condition.

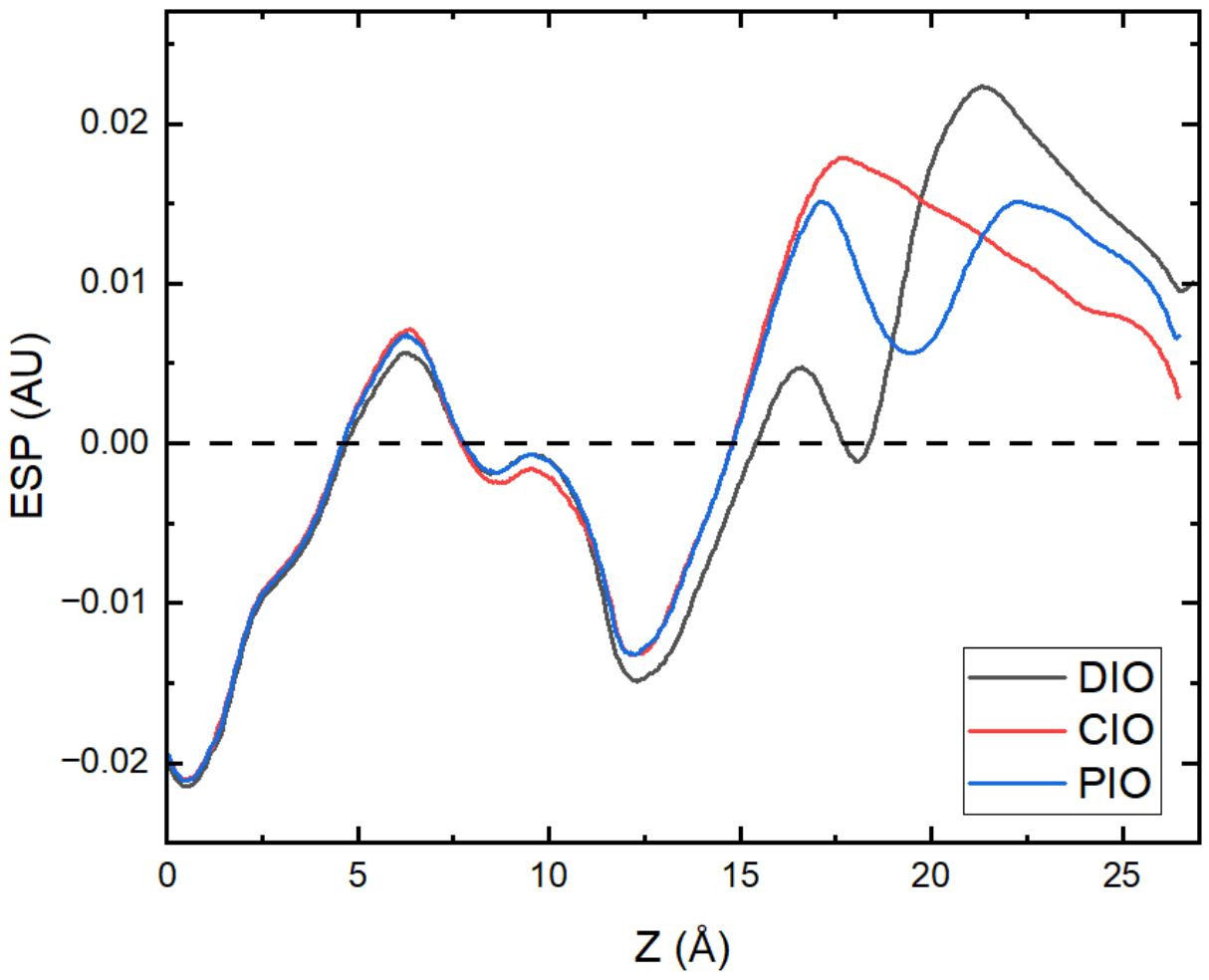

**Fig. S5.** Electrostatic potential along the z-axis of the molecules a) **DIO**, b) **CIO** and c) **PIO**. AU stands for atomic units. d) the same electrostatic potential data plotted overlapped to allow for direct comparison. This data was calculated using a B3LYP-GD3BJ/cc-pVTZ basis set.

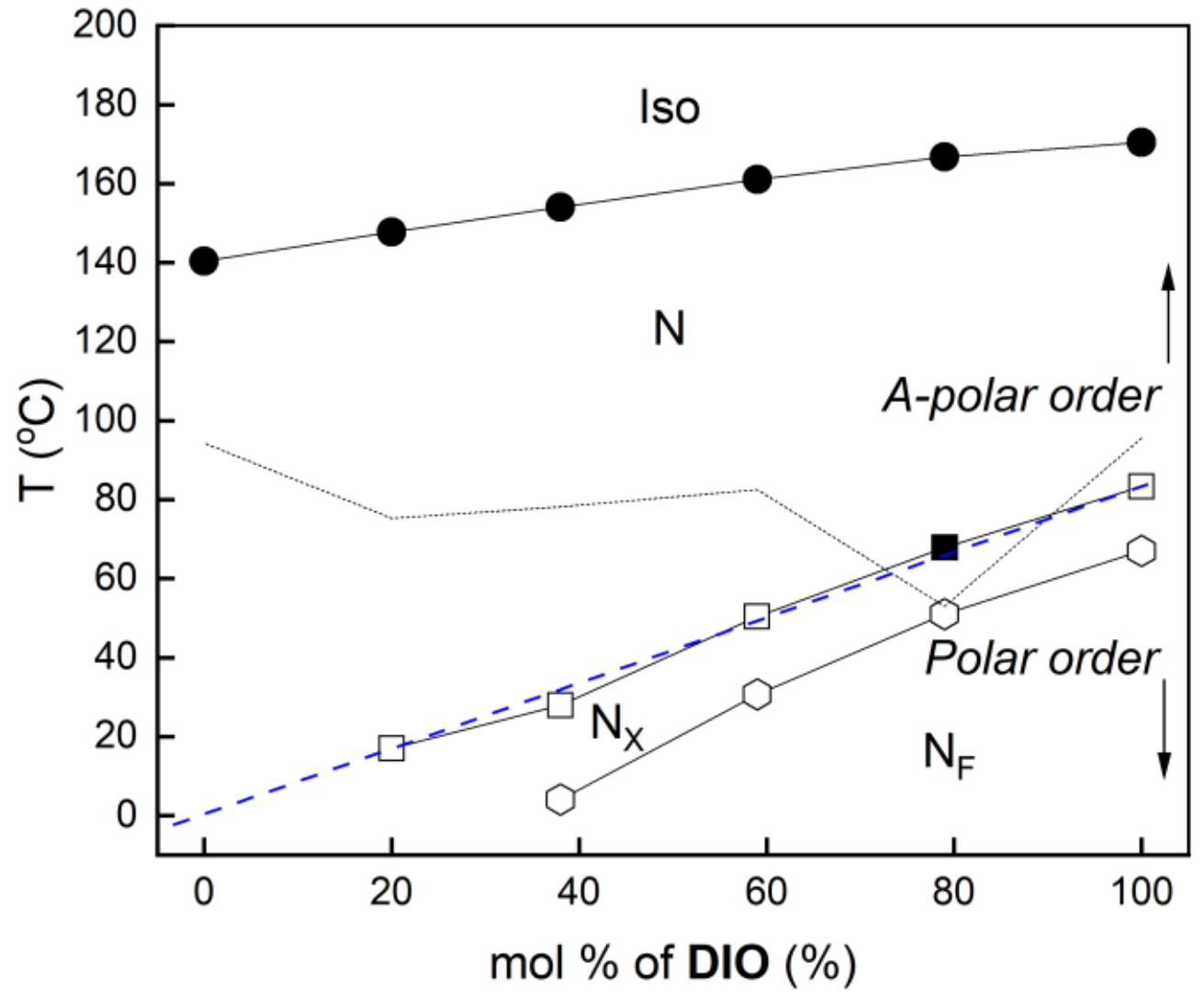

**Fig. S6.** Phase diagram of binary mixtures of **DIO** and compound 9 (**BIO**). Blue dashed line indicates the linear extrapolation of the apolar-polar transition giving a virtual apolar-polar transition temperature of -1.6 °C

## 3. Chemical Synthesis and Structural Characterization

### 3.1 Overall reaction scheme

We synthesised 2’,3,4,5-tetrafluoro-4’-hydroxybiphenyl *via* the Suzuki-Miyaura cross-coupling of 4-bromo-3-fluorophenol with 3,4,5-trifluorobenzene boronic acid with $Pd(OAc)_2$/SPHOS as the catalyst, affording the title compound in 94% yield on ~ 100 mmol scale (28 g). Subsequent esterification (using EDC.HCl and DMAP) with a selection of carboxylic acids, some of which were synthesised as described below (***i1-i6***) and others which were available in house ***(i7-i11),*** afforded the target DIO-homologues detailed in Table 1. For in-house prepared carboxylic acids, iron catalysed Kumada cross coupling of propyl magnesium bromide with either methyl 4-chloro-2-fluorobenzoate or methyl 4-chloro-2,6-difluorobenzoate, followed by basic hydrolysis and acid workup, afforded ***i1*** and ***i12;*** lithiation/carboxylation of *trans* 5-(4-propylcyclohexyl)-1,3-difluorobenzene afforded ***i3***; condensation of 4-borono-2,6-difluorobenzoic acid with 2-propylpropan-1,3-diol in THF with 4A molecular sieves afforded ***i4***; Suzuki-Miyaura cross coupling of 4-bromopropylbenzene with either 3-fluorobenzene boronic acid or 3,5-difluorobenzene boronic acid, followed by lithiation/carboxlyation and acidic workup, afforded ***i5*** and ***i6***.

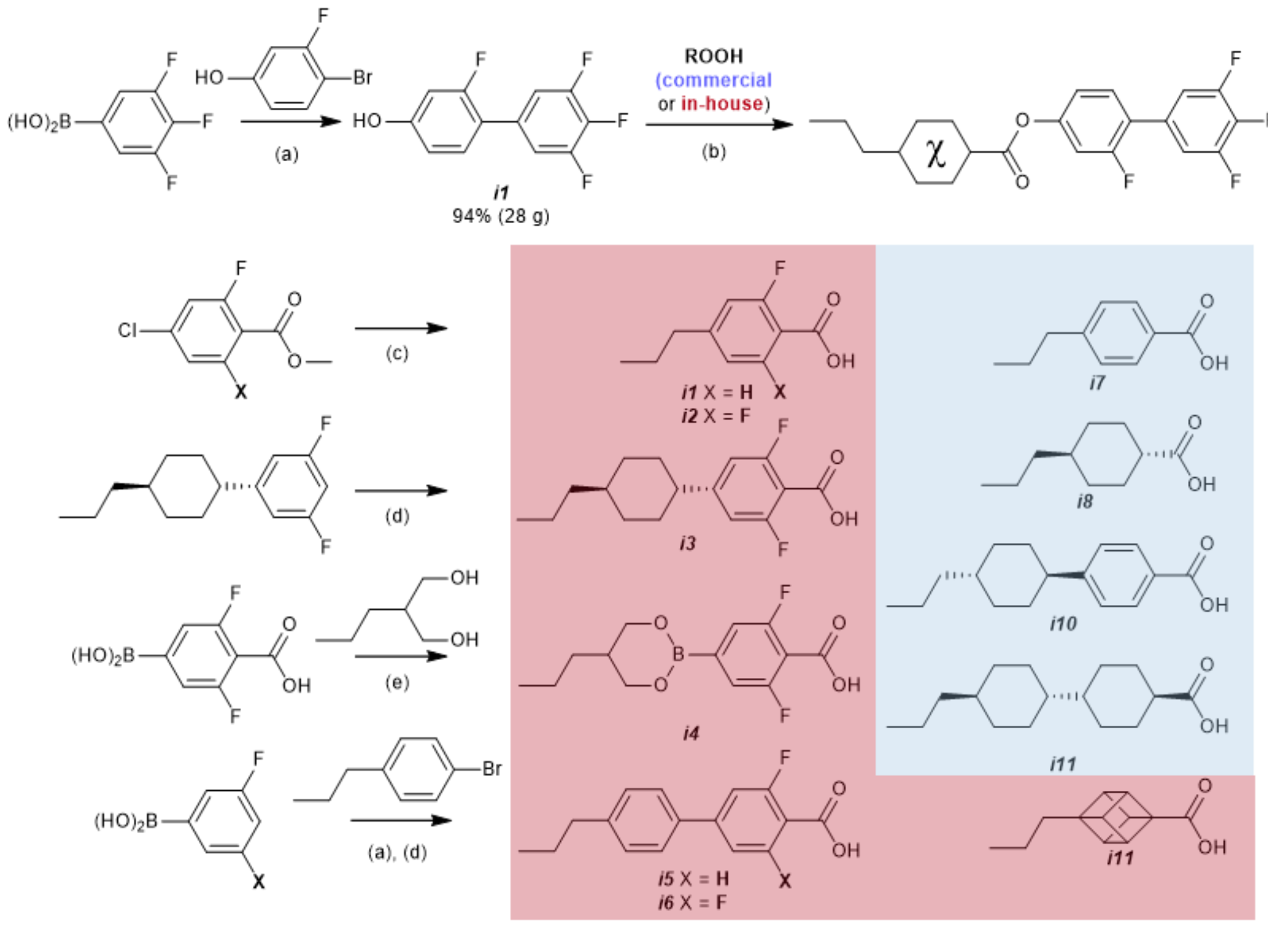

(a) $Pd(OAc)_2$, SPHOS, $K_2CO_3$, THF / $H_2O$, $N_2$, 60 °C;
(b) EDC.HCl, DMAP, DCM;
(c) i) PrMgCl, $Fe(acac)_3$, THF / NMP, $N_2$; ii) KOH, $H_2O$/EtOH; iii) HCl, $H_2O$;
(d) i) nBuLi, THF, -78 °C, $N_2$; ii) $CO_2$, - 78 °C; iii) HCl, $H_2O$, 0 °C;
(e) THF, 4Å MS, 50°C, $N_2$;

**Scheme S1.** Synthetic scheme for synthesis of compounds **1-11** described in this work. Carboxylic acids were either obtained commercially (blue) or synthesised in house (red).

### 3.2 Synthesis of chemical intermediates (*i1-i8*)

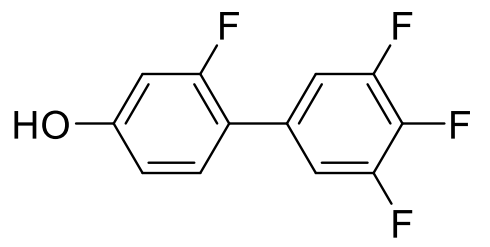

***2',3,4,5-tetrafluoro-4'-hydroxybiphenyl (i1)***

A solution of 4-bromo-2-fluorophenol (23.5 g, 0.123 mol) in biphasic mixture of THF (125 ml) and 2M aqueous $K_2CO_3$ (100 ml) was degassed by sparging with argon for 15 minutes. The Pd-SPHOS catalyst was prepared by degassing 5 ml of THF with argon sparging for 5 minutes, adding solid $Pd(Oac)_2$ (50 mg) and SPHOS (100 mg) and stirring for 5 minutes while sparging with argon. The biphasic reaction mixture was heated to reflux under an argon atmosphere. 3,4,5-Trifluorobenzeneboronic acid (25 g, 0.144 mol) was added as a solid in one portion, followed by the Pd-SPHOS catalyst as a solution in THF. The reaction mixture was heated under reflux with vigorous stirring, under an atmosphere of dry argon, for 1 hour at which point TLC analysis showed total consumption of the starting phenol. The solution was cooled, the aqueous layer separated and washed with ethyl acetate (3x 50 ml), and discarded. The combined organics were sequentially washed with saturated aqueous ammonium chloride (50 ml), brine (50 ml). The organics were then dried over $MgSO_4$, filtered, and volatiles removed *in vacuo*. The crude material was filtered over a short plug of silica gel, eluting with DCM, and then recrystalised from ethanol, affording the title compound as an amorphous white solid. Spectral data matched an authentic sampled purchased from Manchester Organics.

| | |
|---|---|
| Yield: | 28.0 g (94%; amorphous white solid) |
| Rf (DCM): | 0.22 |
| $^1$H NMR (400 MHz): | 7.23-7.30 (1H, m, Ar-**H**), 7.10-7.20 (2H, m, Ar-**H**), 6.68 – 6.75 (2H, m, Ar-**H**), 5.31 (1H, s, ArO**H**). |
| $^{19}$F NMR (376 MHz): | -115.37 (1F, t, $J_{H-F}$= 10.3 Hz, Ar-**F**), -134.77 (2F, *dd*, $J_{F-F}$ = 20.6 Hz, $J_{H-F}$ = 9.0 Hz, Ar-**F**), -162.48 (1F, tt, $J_{F-F}$ = 20.6 Hz, $J_{H-F}$ = 6.5 Hz, Ar-**F**). |

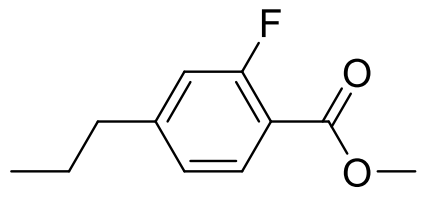

***2-fluoro-4-propyl propylbenzoate (i2)***

Methyl 4-chloro-2-fluorobenzoate (3 g, 16.8mmol) and $Fe(acac)_3$ (352 mg, 1 mmol) were dissolved into in THF (40 ml) and NMP (4 ml) with stirring, under an atmosphere of dry nitrogen gas. Once homogenous, propyl magnesium bromide (1 M, 25ml, 25 mmol) in 2-MeTHF was added dropwise, prompting a colour change from red to green then brown and finally black. The solution was stirred for approximately 1 hour, at which point the colour returned to red/brown. TLC showed consumption of the starting material (Rf = 0.88, DCM) and the formation of a new spot (Rf = 0.84, DCM). The reaction was quenched with saturated aqueous $NHCl_4$ (50 ml). The organic layer was separated and retained, the aqueous was washed with ethyl acetate (3x 20 ml) and discarded. The organic layer was washed with water (5x 50 ml) and finally brine (1x 100 ml). The organic layer was then dried over $MgSO_4$, concentrated to dryness, and trace NMP removed with high vacuum.

Purification with flash chromatography over silica (RediSep 40 g) with a gradient of hexane/EtOAc afforded methyl 2-fluoro-4-propylbenzoate as a colourless oil.

Yield: 2.9 g (94 %; colourless oil)

$R_f$ (DCM): 0.84

1H NMR (400 MHz): 7.83 (1H, t, J = 7.8 Hz, Ar-**H**), 6.99 (1H, dd, J = 7.7 Hz, J = 2.0 Hz, Ar-**H**), 6.93 (1H, dd, J = 11.9 Hz, J = 1.0 Hz, Ar-**H**), 3.90 (3H, s, COOC-$\mathbf{H_3}$), 2.60 (2H, t, J = 7.0 Hz, Ar-C$\mathbf{H_2}$-$CH_2$), 1.63 (2H, m, $CH_2$-C$\mathbf{H_2}$-$CH_3$), 0.92 (3H, t, J = 7.0 Hz, $CH_2$-C$\mathbf{H_3}$)

$^{19}$F NMR (376 MHz): 110.23 (1F, dd, *J* = 7.7 Hz, *J* = 11.9 Hz, Ar-**F**)

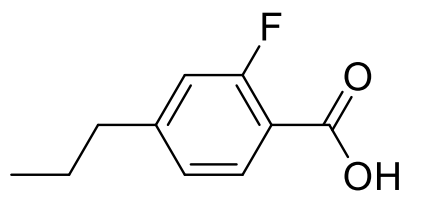

***2-Fluoro-4-propylbenzoic acid (i3)***

Methyl 2-fluoro-4-propylbenzoate (***i1*** , 2.9 g, 16.6 mmol) was dissolved into ethanol (50 ml) and heated under reflux with stirring. A solution of aqueous 2M NaOH (50 ml) was added, and the solution heated under reflux for 18 h. TLC showed complete consumption of the starting material ($R_f$ = 0.84, DCM) and formation of a new spot (Rf = 0.0, DCM). The reaction solution was cooled to ambient temperature and acidified with 2M HCl, affording a white precipitate. The precipitated was collected by filtration, washed with cold water (2x 20 ml), and finally recrystalised from ethanol (-20 °C) to afford the title compound as colourless needles.

Yield 2.5 g (93 %; colourless needles)

1H NMR (400 MHz): 7.96 (1H, t, J = 7.9 Hz, Ar-**H**), 7.07 (1H, ddd, J = 8.1Hz, J = 1.5 Hz, Ar-**H**), 7.01 (1H, dd, J = 12.0 Hz, J = 1.5 Hz, Ar-**H**), 2.67 (2H, t, J = 7.0 Hz, Ar-C$\mathbf{H_2}$-$CH_2$), 1.70 (2H, m, $CH_2$-C$\mathbf{H_2}$-$CH_3$), 0.98 (3H, t, J = 7.4 Hz, $CH_2$-C$\mathbf{H_3}$).

$^{19}$F NMR (376 MHz): -109.04 (1F, dd, *J* = 7.7 Hz, *J* = 12.0 Hz, Ar-**F**)

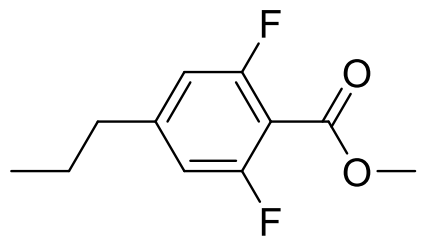

***2,6-Difluoro-4-propyl propylbenzoate (i4)***

Quantities used: methyl 2,6-difluoro-4-chlorobenzoate (1.00 g, 4.84 mmol), propyl magnesium chloride (1M in 2-MeTHF, 6.00 mmol, 6.00 ml), Fe(acac)3 (0.35 g, 1.00 mmol), THF (20.0 ml) NMP (2.00 ml). The experimental procedure was as described for (***i1***). Purification *via* column chromatography with a gradient of hexane/EtOAc over silica afforded the title compound as a colorless oil which was used directly in the synthesis of ***i5*** without further purification or analysis.

Yield: 0.83g, 80%

$R_f$(DCM): 0.52

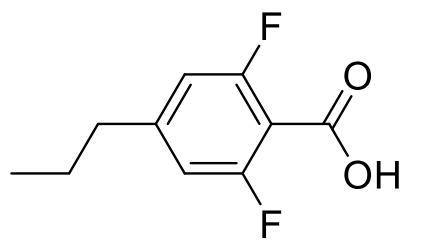

***2,6-Difluoro-4-propylbenzoic acid (i5)***

Quantities used: methyl 2,6-difluoro-4-propylbenzoate (0.83 g, 3.87 mmol), ethanol (20.0 mL), 2M aqueous sodium hydroxide (10 mL). The reaction procedure was as described for the synthesis of ***I-3***, affording the title compound as colourless needles.

Yield: 0.69 g, (89%; colourless needles)

$^{1}$H NMR (400 MHz): 6.73 (d, J = 10.2 Hz, 2H, Ar-**H**), 2.53 (t, J = 7.6 Hz, 2H, Ar-C**H$_2$**-$CH_2$), 1.57 (h, J = 7.5 Hz, 2H, Ch2-C**H$_2$**-$CH_3$), 0.87 (t, J = 7.4 Hz, 3H, $CH_2$-C**H$_3$**).

$^{19}$F NMR (376 MHz): -108.74 (d, *J* = 10.7 Hz, Ar-**F**).

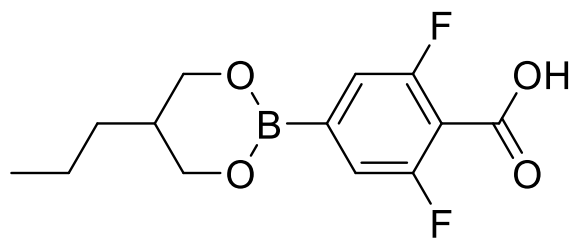

***2-(3,5-difluoro-4-carboxyphenyl)-5-propyl-[1,3,2]dioxaborinane (i6)***

A suspension of 4-borono-2,6-difluorobenzoic acid (1g, 7.58 mmol), 2-propylpropan-1,3-diol (0.91 g, 7.8 mmol) and dried 4Å molecular sieves (1 g) in dry THF (10 ml) was heated with vigorous stirring under an atmosphere of dry nitrogen gas for 24 h. TLC showed complete consumption of the starting material ($R_F$ EtOAc = 0.3) and formation of a new spot ($R_F$ EtOAc ~ 0.5). The volatiles were removed under reduced pressure. The crude material was dissolved into acetone (~ 5 ml), cooled to -20 °C, and precipitated *via* trituration with cold (-20 °C) hexane to afford the title compound as an off-white solid.

Yield: 1.7 g (79%; off-white solid)

Rf (EtOAc): 0.5

$^{1}$H NMR (400 MHz): 7.24 (2H, d, *J* = 9.3 Hz, ArH), 4.09 (2H, *J* = 4.4 Hz, *J* = 11.0 Hz, B-[OCH$H_{ax}$CH-CH$H_{ax}$-O]), 3.69 (2H, *J* = 4.4 Hz, *J* = 11.0 Hz, B-[OCH$H_{eq}$-CH-CH$H_{eq}$-O]), 2.04 (1H, ttt, *J* = 2.0 Hz, *J* = 4.9 Hz, *J* = 8.8 Hz, B-[OCH$_2$-CH(-CH$_2$-CH$_2$...)-CH$_2$-O]), 1.26-1.38 (2H, m, CH-CH$_2$-CH$_2$-CH$_3$), 1.15-1.24 (2H, m, CH-CH$_2$-CH$_2$), 0.88 (3H, t, *J* = 7.2 Hz, Ar-CH$_2$-CH$_2$-C**H$_3$**),

$^{19}$F NMR (376 MHz): -110.78 (2F, S, ArF)

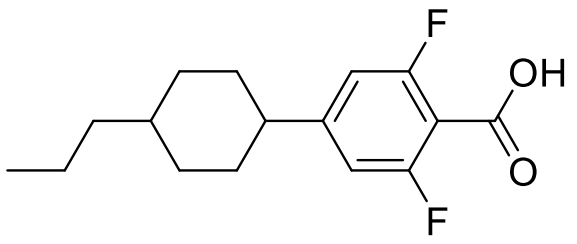

***trans 4-(4-propylcyclohexyl)-2,6-difluorobenzoic acid** (i7)*

A solution of n-butyl lithium (1.6 M in hexanes, 7.50 mL, 12.0 mmol) was added dropwise to a stirred, cooled (-78 °C) solution of *trans* 4-(4-propylcyclohexyl)-2,6-difluorobenzene in anhydrous THF (10 mL) under an atmosphere of dry nitrogen. The aryl lithum solution was allowed to stir for 30 minutes before adding solid carbon dioxide (~ 2.0 g) in a single portion with vigorous stirring and allowing too slowly warm to ambient temperature (~ 2 h). The basic solution was acidified with 2M HCl (~ 50 ml) and extracted with EtOAc. The organics were then dried over $MgSO_4$ and a white solid retrieved under reduced pressure. These were then purified by recrystallization from EtOH/Hexane to give white needles.

Yield: 1.7 g (60%; white needles)

$R_f$ (EtOAc): 0.42

$^1$H NMR (400 MHz): 11.96 (s, 1H, Ar-O**H**), 6.97 – 6.69 (m, 2H, Ar-**H**), 2.59 – 2.37 (m, 1H, Ar-C**H**-($CH_2$)$_2$), 1.98 – 1.80 (m, 4H, CH-C**H₂**-$CH_2$ and Ar-CH(C**H**$_{eq}$)$H_{ax}$ x2)*, 1.47 – 0.96 (m, 10H, $CH_2$(cyclohexane) and $CH_2$-C**H₂**-$CH_3$)*, 0.89 (t, *J* = 7.2 Hz, 3H, $CH_2$-C**H₃**).

$^{19}$F NMR (376 MHz): -108.39 (d, *J* = 10.6 Hz, Ar-F).

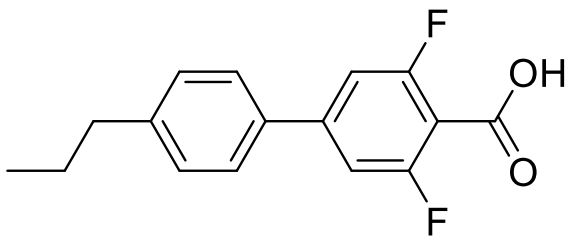

***(4'-propyl-3,5-difluorobiphenyl-4-carboxylic acid) (i8)***

A solution of n-butyl lithium (1.6 M in hexanes, 31.0 mL, 51.0 mmol) was added dropwise to a stirred, cooled (-78 °C) solution of *trans* 4-(4-propylphenyl)-2,6-difluorobenzene (9.79 g, 42.0mmol) in anhydrous THF (10 mL) under an atmosphere of dry nitrogen. The aryl lithium solution was allowed to stir for 30 minutes before adding solid carbon dioxide (~ 2.0 g) in a single portion with vigorous stirring and allowing too slowly warm to ambient temperature (~ 2 h). The basic solution was acidified with 2M HCl (~ 50 ml) and extracted with EtOAc. The organics were then dried over $MgSO_4$ and a white solid retrieved under reduced pressure. These were then purified by recrystallization from EtOH/Hexane to give white needles.

Yield: 9.50 g (81%; white needles)

$R_f$ (EtOAc): 0.37

$^1$H NMR (400 MHz): 7.53 (ddd, J = 8.2, 2.3, 1.8 Hz, 2H, Ar-**H**), 7.32 (ddd, J = 8.2, 1.6, 1.4 Hz, 2H, Ar-**H**), 7.28 – 7.18 (m, 2H, Ar-**H**), 2.67 (t, J = 7.3 Hz, 2H, Ar-C**H₂**-$CH_2$), 1.71 (h, J = 7.5 Hz, 2H, $CH_2$-C**H₂**-$CH_3$), 1.00 (t, J = 7.3 Hz, 3H, $CH_2$-C**H₃**).

$^{19}$F NMR (376 MHz): -107.65 (d, *J* = 10.6 Hz, Ar-**F**).

## 3.3 Characterisation data for final LC compounds

### 3.3.1 General Steglich Esterification Procedure

Unless otherwise noted, a round bottomed flask or 14 ml vial was charged with carboxylic acid (1 mmol, 1 eqv.), phenol (1 mmol, 1 eqv.), EDC.HCL (1.5 mmol, 1.5 eqv.) and DMAP (< 5 mg). A stirrer bar was added, and sufficient DCM added to ensure complete solvation (4 – 50 ml). The reaction vessel was closed with a stopper or cap and stirred until complete consumption of either the acid or phenol as judged by TLC analysis. Once complete, the reaction solution was concentrated and purified by flash chromatography over silica gel with a gradient of hexane/EtOAc using a Combiflash NextGen300+ system. The chromatographed material was dissolved into the minimum quantity of DCM, filtered through a 0.2 micron PTFE filter, concentrated to dryness and finally recrystalised from the indicated solvent system.

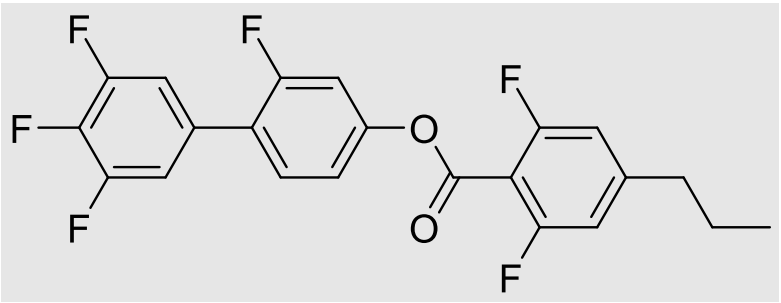

(**1**)

| | |
|---|---|
| Yield: | 326 mg (77 %, colourless crystals); |
| $R_f$ (DCM): | 0.89 |
| $^1$H NMR (400 MHz): | 7.43 (td, *J* = 8.7, 4.3 Hz, 1H, Ar-**H**), 7.23 – 7.07 (m, 6H, Ar-**H**), 6.86 (d, *J* = 10.1 Hz, 2H, Ar-**H**), 2.65 (t, *J* = 7.6 Hz, 2H, Ar-C**H$_2$**-CH$_2$), 1.68 (h, *J* = 7.4 Hz, 2H, CH$_2$-C**H$_2$**-CH$_3$), 0.97 (t, *J* = 7.3 Hz, 3H, CH$_2$-C**H$_3$**). |
| $^{13}$C{$^1$H} NMR (101 MHz): | 164.76, 160.66, 158.17, 152.49, 151.92, 151.82, 149.96, 149.68, 140.73, 138.22, 131.02, 130.57, 130.53, 130.36, 128.87, 126.32, 123.82, 123.69, 118.36, 118.32, 113.32, 113.28, 113.10, 113.06, 110.92, 110.66, 38.15, 24.25, 13.75. |
| $^{19}$F NMR (376 MHz): | -106.69 (d, *J* = 9.0 Hz, Ar-**F**), -109.44 (d, *J* = 10.3 Hz, Ar-**F**), -114.07 – -115.90 (m, Ar-**F**), -134.14 (ddd, *J* = 40.1, 20.7, 8.9 Hz, Ar-**F**), -161.21 (tt, *J* = 20.5, 6.7 Hz, Ar-**F**). |

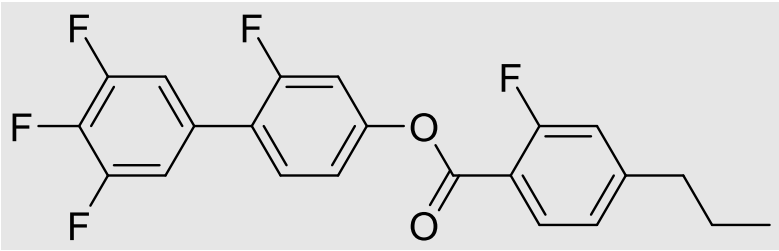

(**2**)

Yield: 332 mg (82 %, colourless crystals);

$R_f$ (DCM): 0.84

$^1$H NMR (400 MHz): 8.01 (t, *J* = 7.8 Hz, 1H, Ar-**H**), 7.42 (t, *J* = 8.7 Hz, 1H, Ar-**H**), 7.25 – 6.98 (m, 6H, Ar-**H**), 2.68 (t, *J* = 6.8 Hz, 2H, Ar-C**H**$_2$-CH-$_2$), 1.70 (h, *J* = 7.4 Hz, 2H, $CH_2$-C**H**$_2$-$CH_3$), 0.98 (t, *J* = 7.3 Hz, 3H, $CH_2$-C**H**$_3$).

$^{13}$C{$^1$H} NMR (101 MHz): 163.81, 162.31, 162.27, 161.21, 160.63, 158.14, 152.45, 152.35, 152.27, 151.52, 151.41, 150.01, 149.90, 140.75, 138.23, 132.41, 130.97, 130.58, 130.54, 124.53, 124.50, 123.96, 123.84, 118.32, 118.28, 117.18, 116.96, 114.66, 114.57, 113.33, 113.29, 113.24, 113.11, 113.07, 110.88, 110.63, 37.86, 23.88, 13.66.

$^{19}$F NMR (376 MHz): -108.31 (dd, *J* = 11.9, 7.5 Hz, Ar-**F**), -114.66 (t, *J* = 9.9 Hz, Ar-**F**), -134.25 (dd, *J* = 20.6, 8.8 Hz, Ar-**F**), -161.31 (tt, *J* = 20.7, 6.6 Hz), Ar-**F**.

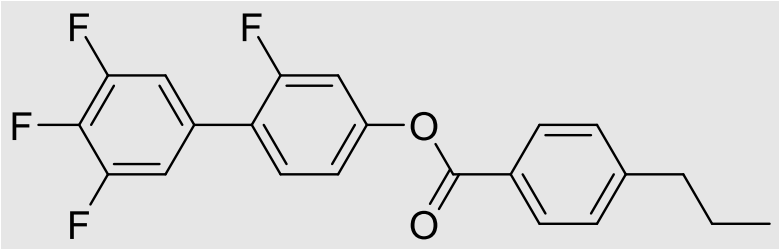

(**3**)

Yield: 0.28 mg (72 %, colourless crystals);

$R_f$ (DCM): 0.83

$^1$H NMR (400 MHz): 8.11 (ddd, *J* = 8.3, 1.9, 1.8 Hz, 2H, Ar-**H**), 7.43 (t, *J* = 8.6 Hz, 1H, Ar-**H**), 7.34 (ddd, *J* = 8.5, 1.9, 1.8 Hz, 2H, Ar-**H**), 7.23 – 7.16 (m, 2H, Ar-**H**), 7.14 (t, *J* = 2.8 Hz, 1H, Ar-**H**), 7.13 – 7.09 (m, 1H, Ar-**H**), 2.70 (t, *J* = 6.8 Hz, 2H, Ar-C**H**$_2$-$CH_2$), 1.70 (h, *J* = 7.5 Hz, 2H, $CH_2$-C**H**$_2$-$CH_3$), 0.98 (t, *J* = 7.4 Hz, 3H, $CH_2$-C**H**$_3$).

$^{13}$C{$^1$H} NMR (101 MHz): 164.76, 160.66, 158.17, 152.49, 151.92, 151.82, 149.96, 149.68, 140.73, 138.22, 131.02, 130.57, 130.53, 130.36, 128.87, 126.32, 123.82, 123.69, 118.36, 118.32, 113.32, 113.28, 113.10, 113.06, 110.92, 110.66, 77.35, 38.15, 24.25, 13.75.

$^{19}$F NMR (376 MHz): -114.72 (t, *J* = 9.7 Hz, Ar-**F**), -134.26 (dd, *J* = 20.5, 8.8 Hz, Ar-**F**), -161.35 (tt, *J* = 20.7, 6.6 Hz, Ar-**F**).

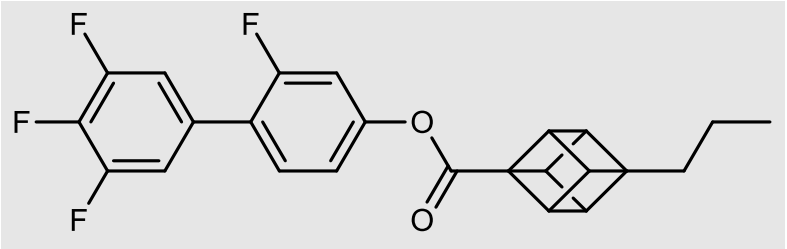

**(4)**

Yield: 152 mg (78 %; colourless prisms)

$R_f$ (DCM): 0.91

$^1$H NMR (400MHz): 7.36-7.26 (1H, m, Ar**H**), 7.14-7.04 (2H, m, Ar**H**), 6.99-6.91 (2H, m, Ar**H**), 4.22-4.14 (3H, m, Cub**H**), 3.78-3.72 (3H, m, Cub**H**), 1.57-1.47 (2H, m, Cub-C**H**$_2$-CH$_2$-CH$_3$), 1.36-1.23 (2H, m, Cub-CH$_2$-C**H**$_2$-CH$_3$), 0.88 (3H, t, $J$ = 7.4 Hz, Cub-CH$_2$-CH$_2$-C**H**$_3$)

$^{13}$C{$^1$H} NMR (101MHz): 170.37, 160.60, 158.10, 152.6-152.2 (m), 151.66 (m), 140.41 (m), 130.44 (d, $J$ = 4.1 Hz), 123.31 (d, $J$ = 15.4 Hz), 118.18 (d, $J$ = 3.7 Hz), 113.61-112.87 (m), 110.59 (d, $J$ = 25.7 Hz), 60.01, 56.13, 46.38, 46.30, 37.41, 17.53, 14.20

$^{19}$F NMR (376 MHz): -114.92 (1F, t, $J$ = 9.9 Hz, Ar**F**), -134.36 (2F, dd, $J$ = 8.7 Hz, $J$ = 20.5 Hz, Ar**F**), -161.47 (1F, tt, $J$ = 6.5 Hz, $J$ = 20.5 Hz, Ar**F**)

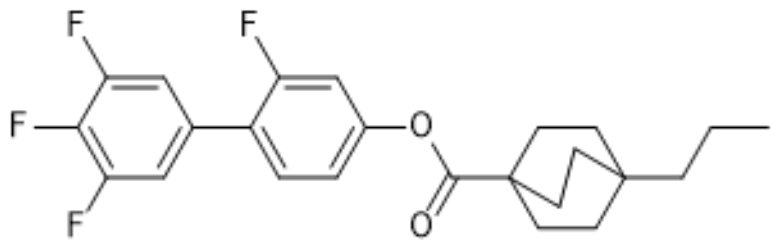

**(5)**

| | |
|---|---|
| Yield: | 161 mg (77%, colurless needles) |
| $R_{f\,(DCM)}$: | 0.95 |
| $^{1}$H NMR (400 MHz): | 7.32-7.24 (1H, m, Ar$\underline{H}$), 7.14-7.01 (2H, m, Ar$\underline{H}$), 6.90-6.82 (2H, m, Ar$\underline{H}$), 1.89-1.81 (6H, m, BCO$\underline{H}$), 1.43-1.35 (6H, m, BCO$\underline{H}$), 1.25-1.08 (2H, m, -C$\underline{H}_2$-$CH_2$-$CH_3$), 1.08-0.98 (2H, m, -$CH_2$-C$\underline{H}_2$-$CH_3$), 0.81 (3H, t, $J$ = 7.7 Hz, (2H, m, -$CH_2$-$CH_2$-C$\underline{H}_3$) |
| $^{13}$C{$^{1}$H} NMR (101 MHz): | 176.29, 160.58, 159.09, 151.91 (d, $J$ =12.0 Hz), 151.14 (dd, $J$ = 250.0 Hz, $J$ = 9.8 Jz), 140.63 (m), 138.31 (m), 130.40 (d, $J$ = 4.0 Hz), 118.12 (d, $J$ = 3.6 Hz), 113.40-112.70 (m), 110.55 (d, $J$ = 25.6 Hz), 49.31, 39.50, 30.57, 30.30, 28.60, 16.92, 15.04 |
| $^{19}$F NMR (376 MHz): | -114.99 (1F, t, $J$ = 9.9 Hz, Ar$\underline{F}$), -134.33 (2F, dd, $J$ = 20.6 Hz, $J$ = 8.8 Hz, Ar$\underline{F}$), -161.44 (1F, tt, $J$ = 20.6 Hz, $J$ = 6.6 Hz) |

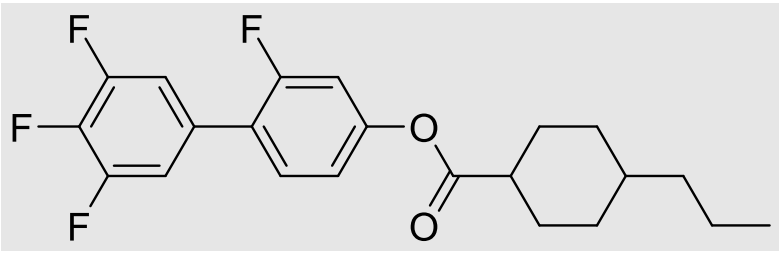

(**6**)

| | |
|---|---|
| Yield: | 350 mg (90 %, colourless crystals); |
| $R_f$ (DCM): | 0.90 |
| $^{1}$H NMR (400 MHz): | 7.27 (t, $J$ = 8.7 Hz, 1H, Ar-**H**), 7.06 (ddd, $J$ = 8.8, 6.5, 1.3 Hz, 2H, Ar-**H**), 6.91 – 6.84 (m, 2H, Ar-**H**), 2.40 (tt, $J$ = 12.2, 3.6 Hz, 1H, Ar-C**H**-$(CH_2)_2$), 2.11 – 1.96 (m, 2H, CH-[(C$\mathbf{H_{eq}}$)$H_{ax}$]x2-$CH_2$), 1.85 – 1.73 (m, 2H, CH-[(C$\mathbf{H_{ax}}$)$H_{eq}$]x2-$CH_2$), 1.56 – 1.39 (m, 2H $CH_2$-[(C$\mathbf{H_{eq}}$)$H_{ax}$]x2-CH), 1.32 – 1.07 (m, 5H $CH_2$-[(C$\mathbf{H_{ax}}$)$H_{eq}$]x2-CH and $CH_2$CH-C$\mathbf{H_2}$-$CH_2$), 1.01 – 0.83 (m, 3H, $CH_2$-C**H**-$CH_2$ and $CH_2$-C$\mathbf{H_2}$-$CH_3$), 0.81 (t, $J$ = 7.2 Hz, 3H, $CH_2$-C$\mathbf{H_3}$). |
| $^{13}$C{$^{1}$H} NMR (101 MHz): | 174.18, 160.55, 159.86, 151.76, 151.66, 130.41, 118.09, 113.03, 113.00, 110.64, 110.39, 77.31, 76.99, 76.67, 43.57, 39.38, 36.57, 32.14, 28.93, 19.88, 14.31. |
| $^{19}$F NMR (376 MHz): | -114.90 (t, $J$ = 9.7 Hz, Ar-**F**), -134.31 (dd, $J$ = 20.6, 8.8 Hz, Ar-**F**), -161.43 (tt, $J$ = 20.6, 6.6 Hz, Ar-**F**). |

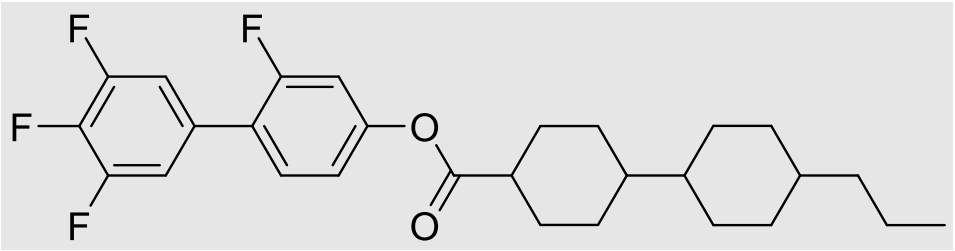

(**7**)

Yield: 285 mg (60 %, colourless crystals);

$R_f$ (DCM): 0.84

$^1$H NMR (400 MHz): 7.36 (t, *J* = 8.7 Hz, 1H, Ar-**H**), 7.20 – 7.10 (m, 2H, Ar-**H**), 7.01 – 6.92 (m, 2H, Ar-**H**), 2.47 (tt, *J* = 12.2, 3.5 Hz, 1H, Ar-C**H**-($CH_2$)$_2$), 2.21 – 2.09 (m, 2H, CH-[(C$\mathbf{H_{eq}}$)$H_{ax}$]x2-$CH_2$), 1.97 – 1.64 (m, 9H, CH-[(C$\mathbf{H_{ax}}$)$H_{eq}$]x2-$CH_2$, CH-[(C$\mathbf{H_{eq}}$)$H_{ax}$]x2-$CH_2$, and ($CH_2$)$_2$-CH-C**H**-($CH_2$)$_2$)$^*$, 1.64 – 1.46 (m, 2H), 1.39 – 1.24 (m, 3H), 1.23 – 0.93 (m, 11H), 0.93 – 0.76 (m, 5H).

$^{13}$C{$^1$H} NMR (101 MHz): 174.20, 160.58, 158.09, 151.69, 130.42, 118.12, 118.09, 113.24, 113.05, 110.67, 110.41, 77.34, 77.02, 76.70, 43.65, 43.21, 42.48, 39.78, 37.59, 33.51, 30.01, 29.23, 29.08, 28.96, 28.76, 20.04, 14.42.

$^{19}$F NMR (376 MHz): -114.88 (t, *J* = 9.6 Hz. Ar-**F**), -134.28 (dd, *J* = 20.4, 8.9 Hz, Ar-**F**), -161.39 (tt, *J* = 20.8, 6.7 Hz, Ar-**F**).

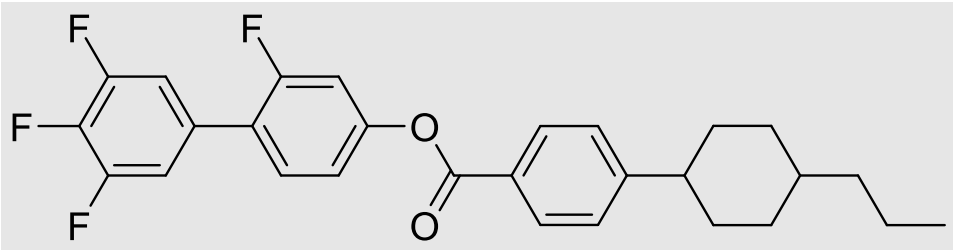

(**8**)

Yield: 300 mg (61 %, colourless crystals);

$R_f$ (DCM): 0.80

$^1$H NMR (400 MHz): 8.12 (ddd, *J* = 8.4, 1.9, 1.8 Hz, 2H, Ar-**H**), 7.42 (t, *J* = 8.6 Hz, 1H, Ar-**H**), 7.37 (ddd, *J* = 8.2, 2.2, 1.8 Hz, 2H, Ar-**H**), 7.22 – 7.15 (m, 2H, Ar-**H**), 7.14 – 7.08 (m, 1H, Ar-**H**), 2.58 (tt, *J* = 12.0, 3.0 Hz, 1H, Ar-C**H**-($CH_2)_2$), 1.91 (m, 5H, ($CH_2)_2$-CH-C**H₂**-$CH_2$ and CH-[(C**H**$_{\mathbf{eq}}$)$H_{ax}$]x2-$CH_2$)*, 1.55 – 1.43 (m, 2H, CH-[(C**H**$_{\mathbf{ax}}$)$H_{eq}$]x2-$CH_2$), 1.43 – 1.28 (m, 2H, $CH_2$-[(C**H**$_{\mathbf{eq}}$)$H_{ax}$]x2-CH), 1.28 – 1.20 (m, 2H, $CH_2$-[(C**H**$_{\mathbf{ax}}$)$H_{eq}$]x2-CH), 1.16 – 1.00 (m, 2H, $CH_2$-C**H**$_{\mathbf{2}}$-$CH_3$), 0.92 (t, *J* = 7.2 Hz, 3H, $CH_2$-C**H**$_{\mathbf{3}}$).

$^{13}$C{$^1$H} NMR (101 MHz): 164.73, 160.66, 158.16, 154.76, 152.42, 151.93, 151.82, 149.97, 140.74, 131.01, 130.57, 130.53, 130.44, 127.28, 126.39, 123.81, 123.68, 118.36, 118.32, 113.32, 113.28, 113.10, 110.91, 110.66, 44.90, 39.65, 36.97, 34.04, 33.38, 20.03, 14.40.

$^{19}$F NMR (376 MHz): -114.73 (t, *J* = 9.8 Hz, Ar-**F**), -134.25 (dd, *J* = 20.5, 8.7 Hz, Ar-**F**), -161.34 (tt, *J* = 20.8, 6.5 Hz, Ar-**F**).

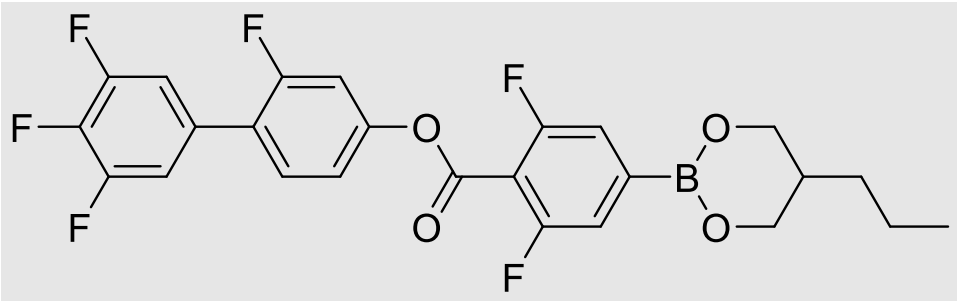

(**9 / BIO**)

| | |
|---|---|
| Yield: | 300mg (66%, colourless crystals); |
| $R_f$ (DCM): | 0.65 |
| $^1$H NMR (400 MHz): | 7.24 (2H, d, *J* = 9.3 Hz, Ar<u>H</u>), 7.36-7.49 (3H, m, Ar<u>H</u>), 4.10-4.30 (2H, m, B-[OCH<u>H</u>$_{ax}$CH-CH<u>H</u>$_{ax}$-O]), 7.14-7.23 (4H, m, Ar<u>H</u>), 3.75-3.83 (2H, m, B-[OCH<u>H</u>$_{eq}$-CH-CH<u>H</u>$_{eq}$-O]), 2.00 – 2.02 (1H, m, B-[OCH$_2$-C<u>H</u>(-CH$_2$-CH$_2$...)-CH$_2$-O]), 1.35-1.46 (2H, m, CH-CH$_2$-C<u>H</u>$_2$-CH$_3$), 1.24-1.32 (2H, m, CH-C<u>H</u>$_2$-CH$_2$), 0.96 (3H, t, *J* = 7.2 Hz, z-CH$_2$-CH$_2$-C**H**$_3$) |
| $^{13}$C{$^1$H} NMR (101 MHz): | 160.62, 160.35 (dd, *J* = 256.7 Hz, *J* = 26.1 Hz), 159.71, 158.12, 140.78 (t, *J* = 16.1 Hz), 138.27 (t, *J* = 15.0 Hz), 130.65 (d, *J* = 3.7 Hz), 124.25 (d, *J* = 12.8 Hz), 118.17 (d, *J* = 3.7 Hz), 116.81 (d, *J* = 3.7 Hz), 113.45-112.79 (m), 110.99 (d, *J* = 17.3 Hz), 110.66 (d, *J* = 26.1 Hz), 67.01 (BO-<u>C</u>-C), 36.15 (-OC**C**HCO-), 30.28 (HC-**C**H$_2$CH$_2$CH$_3$), 19.96 (CH$_2$**C**H$_2$CH$_3$), 14.15 (CH$_2$CH$_2$**C**H$_3$), |
| $^{19}$F NMR (376 MHz): | -110.75 (2F, d, $J_{F-H}$ = 9.5 Hz, Ar<u>F</u>) , -114.44 (1F, t, $J_{F-F}$ = 9.7 Hz, Ar<u>F</u>), -134.26 (2F, $J_{H-F}$ = 8.8 Hz, $J_{F-F}$ = 20.7 Hz, Ar<u>F</u>), -161.29 (2F, dd, $J_{H-F}$ = 6.7 Hz, *J* = 20.7, Ar<u>F</u>) |

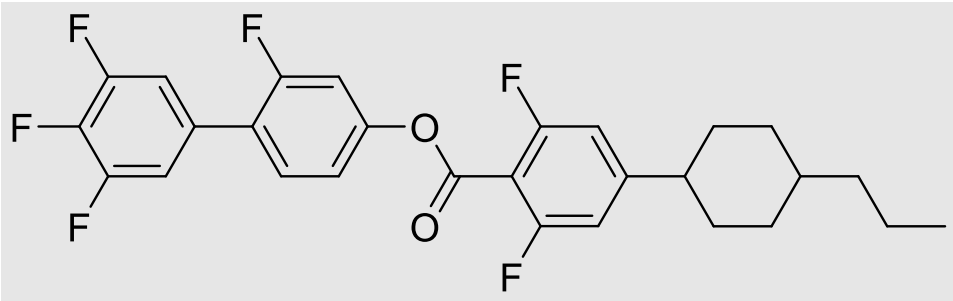

(**10 / ClO**)

Yield: 355 mg (71 %, colourless crystals);

$R_f$ (DCM): 0.82

$^1$H NMR (400 MHz): 7.43 (t, J = 8.6 Hz, 1H, Ar-**H**), 7.22 – 7.10 (m, 5H, Ar-**H**), 6.89 (ddd, J = 10.7, 1.1, 1.1 Hz, 2H, Ar-**H**), 2.64 – 2.46 (m, 1H, Ar-C**H**-$(CH_2)_2$), 1.91 (ddd, J = 13.5, 6.4, 3.3 Hz, 5H, 2x (C$\mathbf{H_{eq}}$)$H_{ax}$-CH-$CH_2$, (CH2)-C**H**-$CH_2$, and CH-C$\mathbf{H_2}$-$CH_2$), 1.52 – 0.95 (m, 9H, 4x C$\mathbf{H_2}_{cyclohexane}$ and CH2-$CH_2$-$CH_3$), 0.91 (t, J = 7.2 Hz, 3H, $CH_2$-C$\mathbf{H_3}$).

$^{13}$C{$^1$H} NMR (101 MHz): 162.59, 162.53, 160.62, 160.02, 159.96, 159.58, 158.12, 156.27, 156.18, 151.15, 151.04, 130.64, 130.60, 124.12, 118.21, 118.17, 113.34, 113.31, 113.12, 113.09, 110.90, 110.87, 110.80, 110.68, 110.65, 110.54, 106.70, 44.64, 39.49, 36.81, 33.64, 33.07, 19.98, 14.36.

$^{19}$F NMR (376 MHz): -109.18 (d, *J* = 10.6 Hz, Ar-**F**), -114.45 (t, *J* = 9.7 Hz, Ar-**F**), -134.19 (dd, *J* = 20.6, 8.7 Hz, Ar-**F**), -161.20 (tt, *J* = 20.4, 6.6 Hz, Ar-**F**).

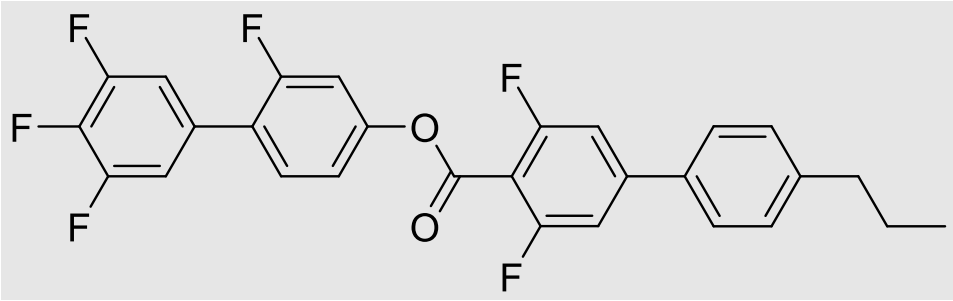

(**11 / PIO**)

Yield: 365 mg (73 %, colourless crystals);

$R_f$ (DCM): 0.89

$^{1}H$ NMR (400 MHz): 7.52 (ddd, *J* = 8.3, 2.0, 1.8 Hz, 2H, Ar-**H**), 7.44 (t, *J* = 8.6 Hz, 1H, Ar-**H**), 7.36 – 7.26 (m, 2H, Ar-**H**), 7.23 – 7.14 (m, 4H, Ar-**H**), 2.66 (t, *J* = 6.8 Hz, 2H, Ar-C**H₂**-CH₂), 1.69 (h, *J* = 7.5 Hz, 2H, CH₂-C**H₂**-CH₃), 0.98 (t, *J* = 7.3 Hz, 3H, CH₂-C**H₃**).

$^{13}C\{^{1}H\}$ NMR (101 MHz): 162.90, 160.64, 160.34, 160.27, 159.45, 158.14, 151.12, 151.01, 147.95, 144.58, 134.85, 130.69, 130.65, 129.41, 126.89, 118.21, 118.17, 113.36, 113.32, 113.14, 110.82, 110.60, 110.56, 110.37, 110.34, 37.72, 24.44, 13.81.

$^{19}F$ NMR (376 MHz): -108.25 (d, *J* = 10.5 Hz, Ar-**F**), -114.36 (t, *J* = 9.8 Hz, Ar-**F**), -134.16 (dd, *J* = 20.7, 8.6 Hz, Ar-**F**), -161.16 (tt, *J* = 20.6, 6.6 Hz, Ar-**F**).

## 3.4 Example NMR Spectra

Fig. S7-S11 are given as representative examples of the NMR spectra of **1-11**. The remaining NMR data is available at the attached DOI (https://doi.org/10.5518/1510).

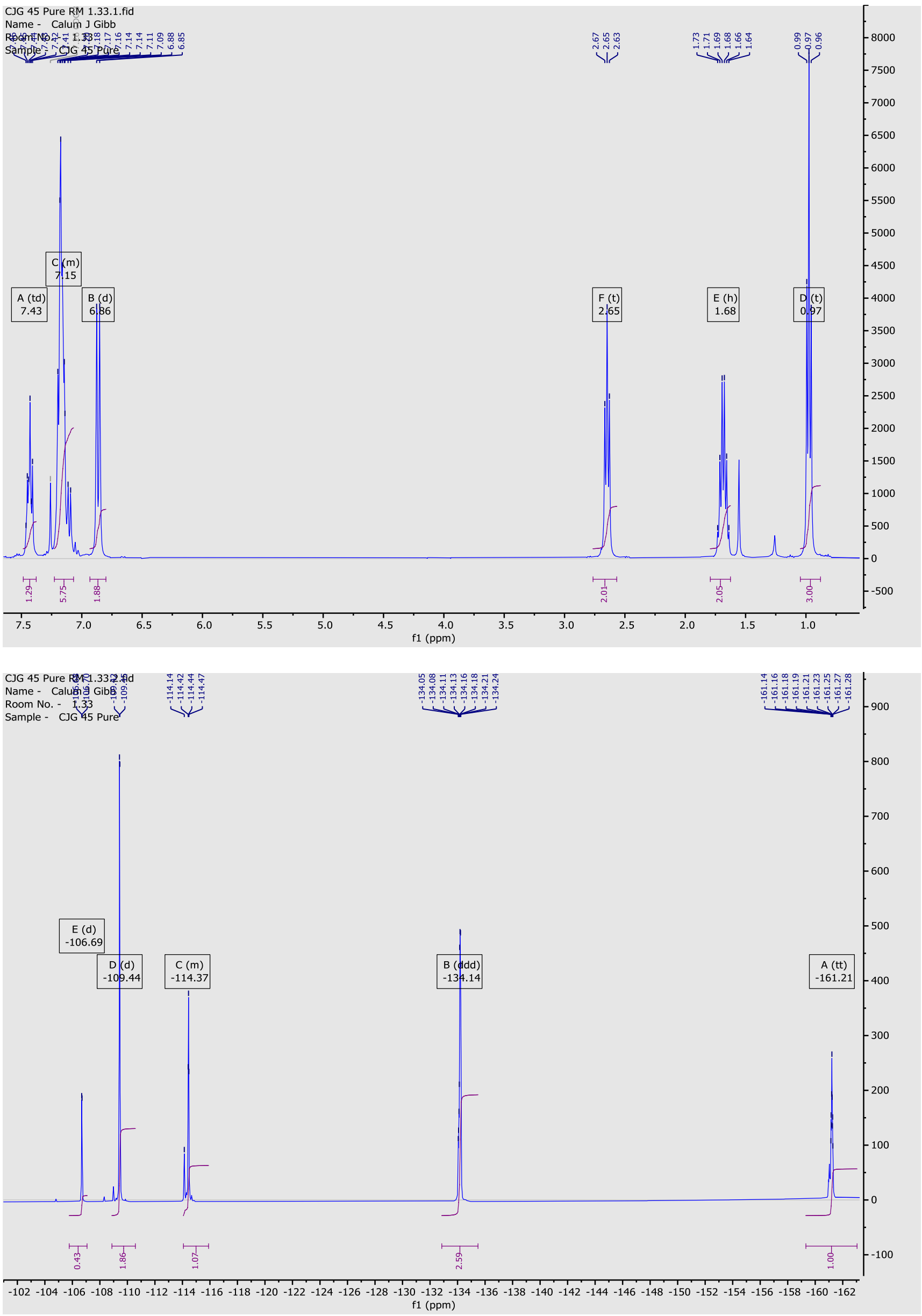

Fig. S7: E NMR spectra of **1**: $^{1}H$ [top], $^{19}F$ [middle], and $^{13}C\{^{1}H\}$ [bottom].

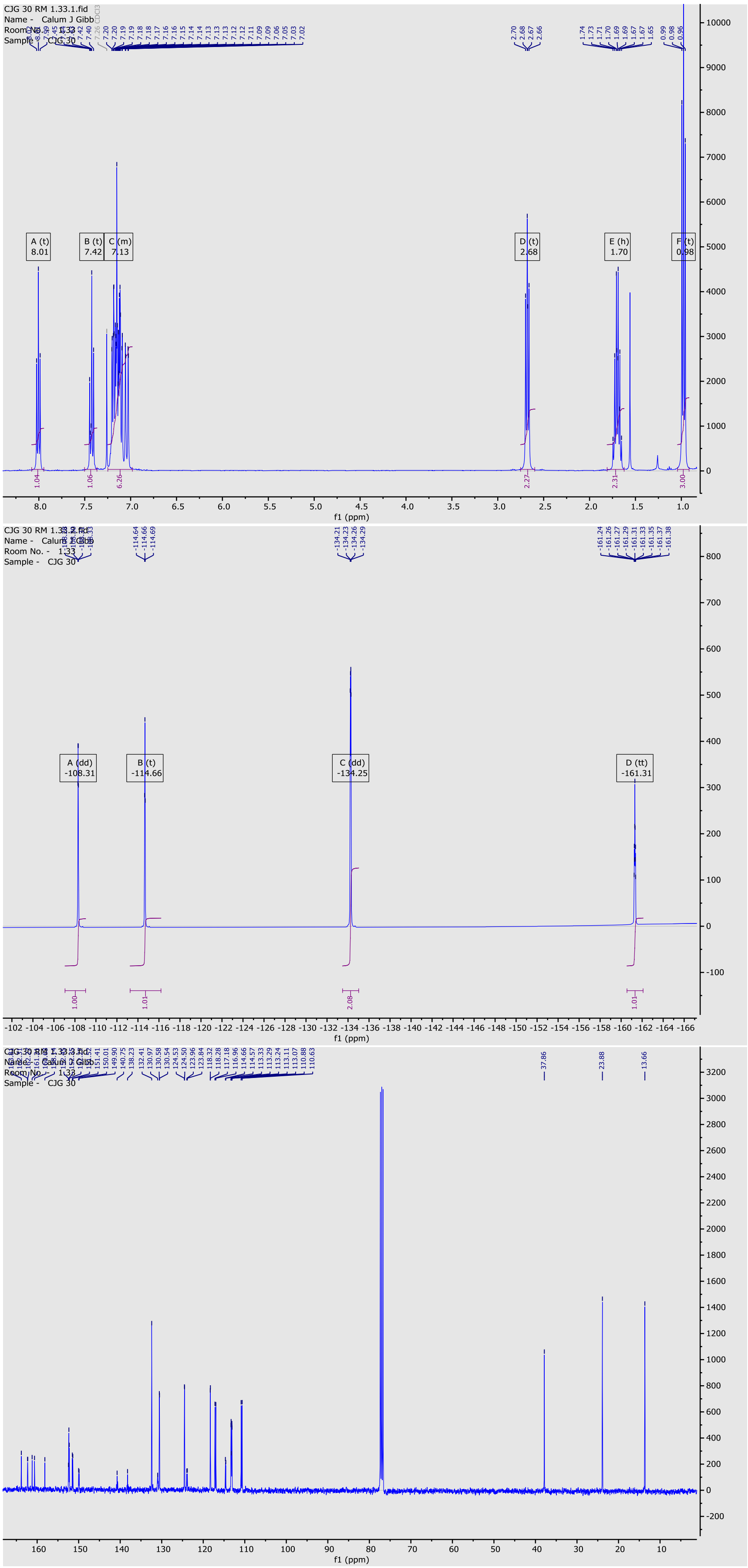

Fig. S8: E NMR spectra of **2**: $^{1}H$ [top], $^{19}F$ [middle], and $^{13}C\{^{1}H\}$ [bottom].

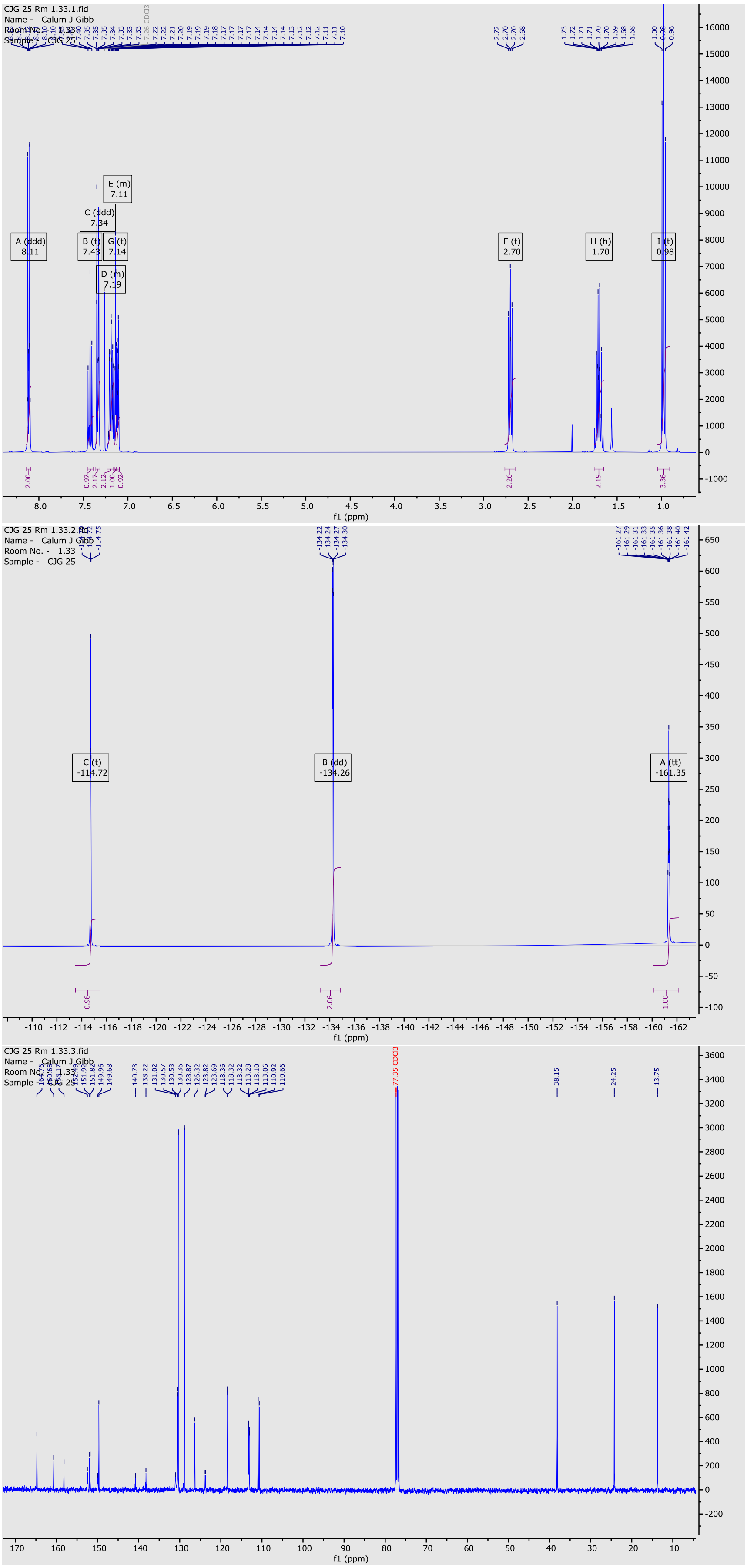

Fig. S9: E NMR spectra of **3**: $^{1}H$ [top], $^{19}F$ [middle], and $^{13}C\{^{1}H\}$ [bottom].

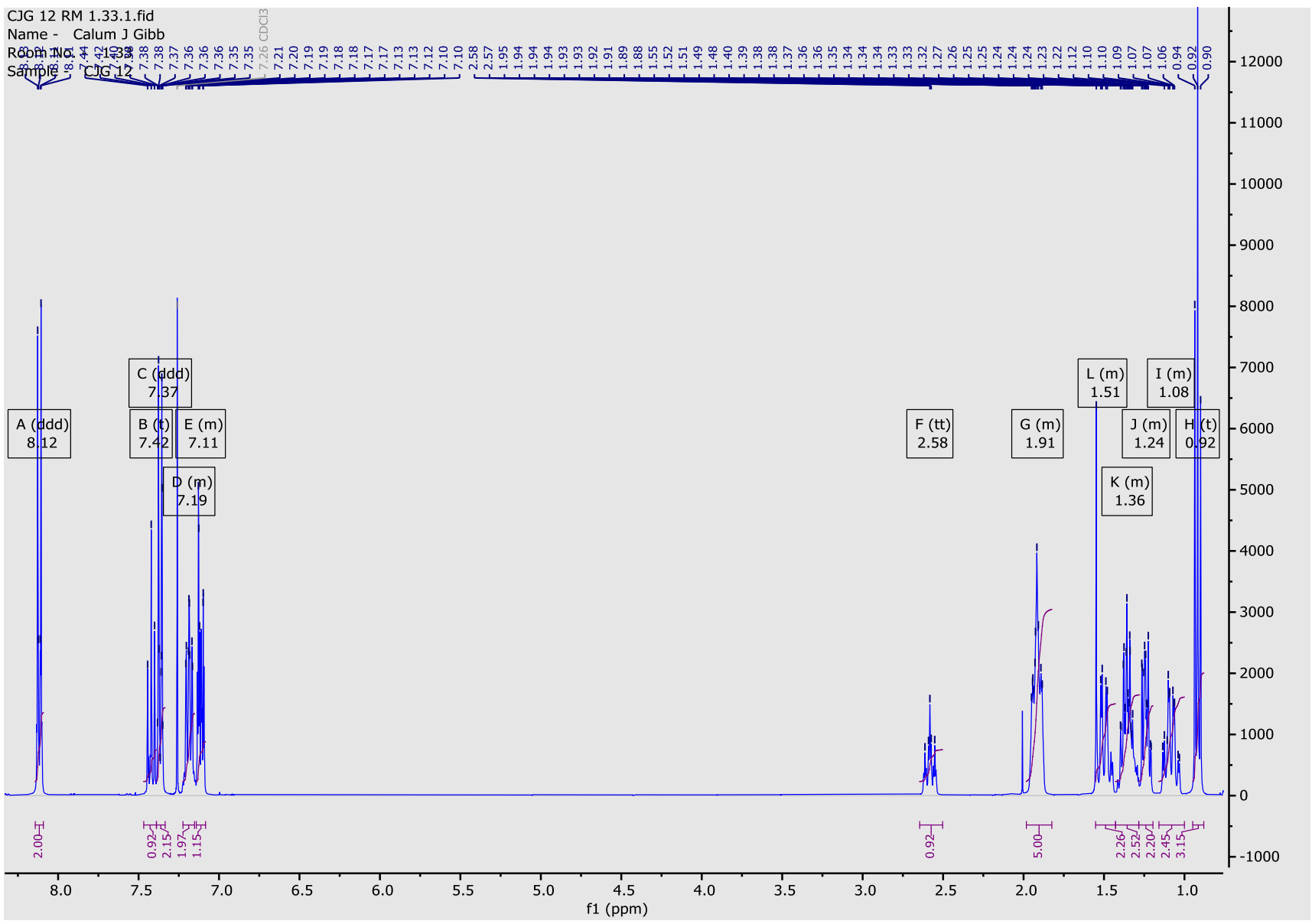

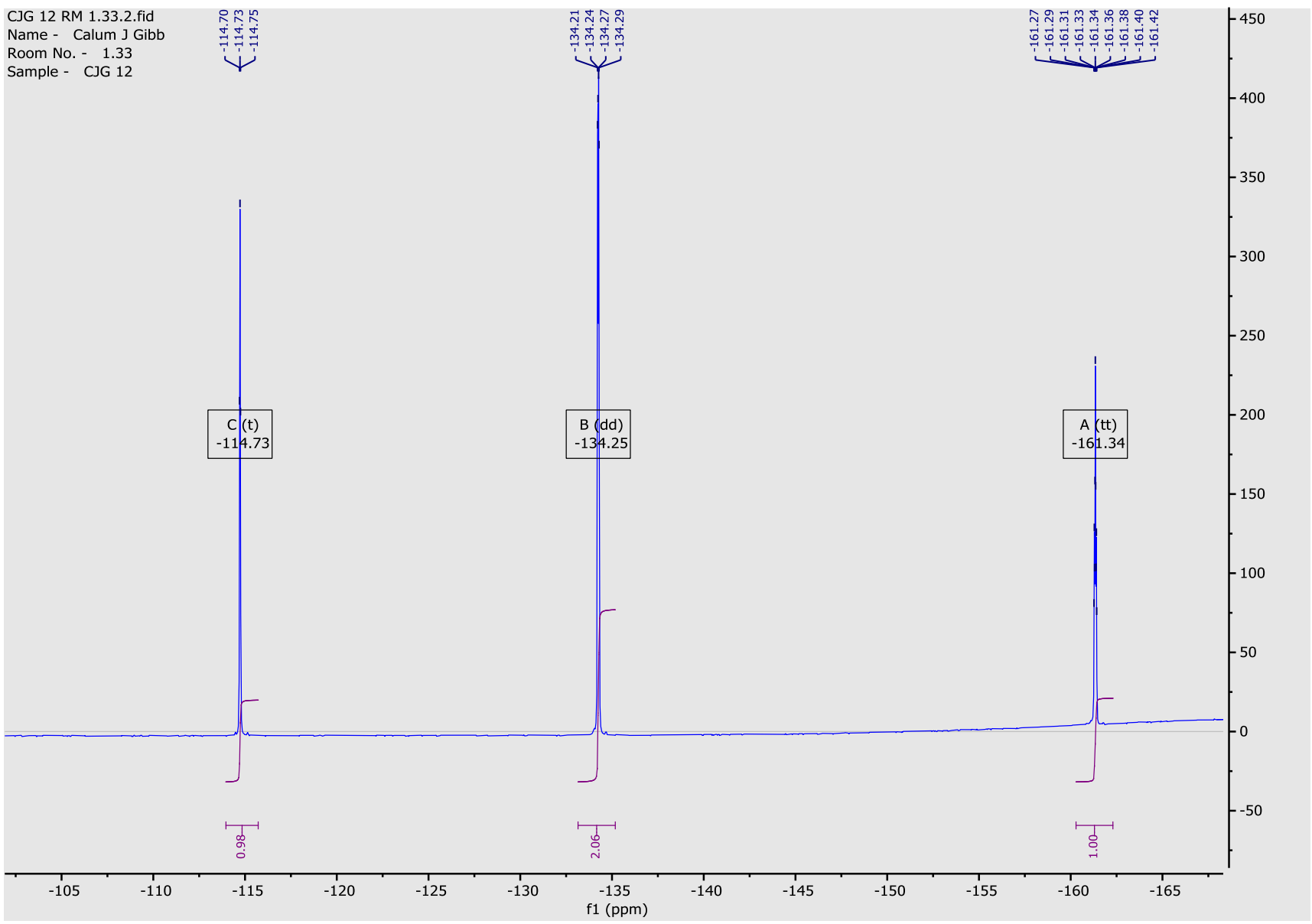

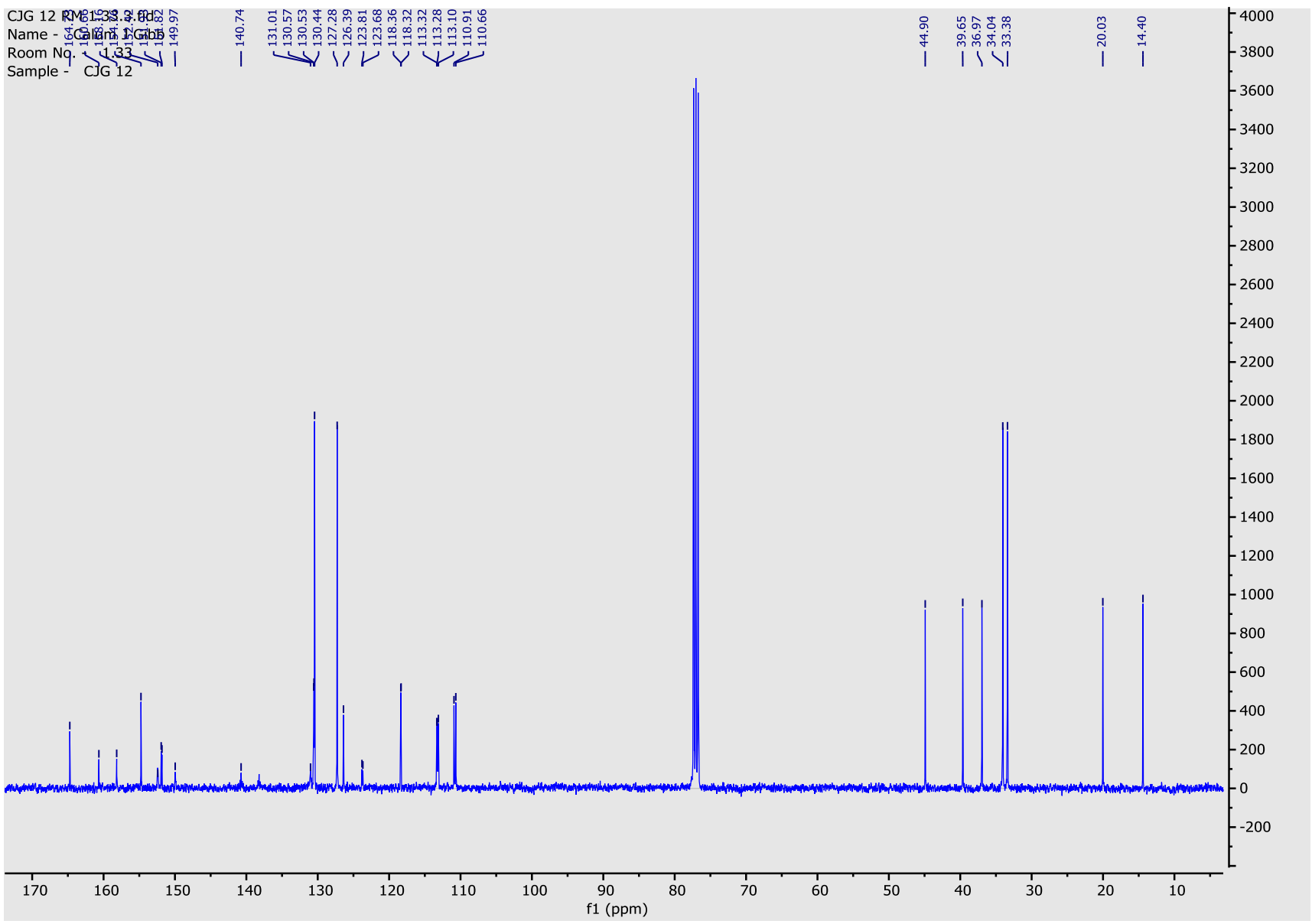

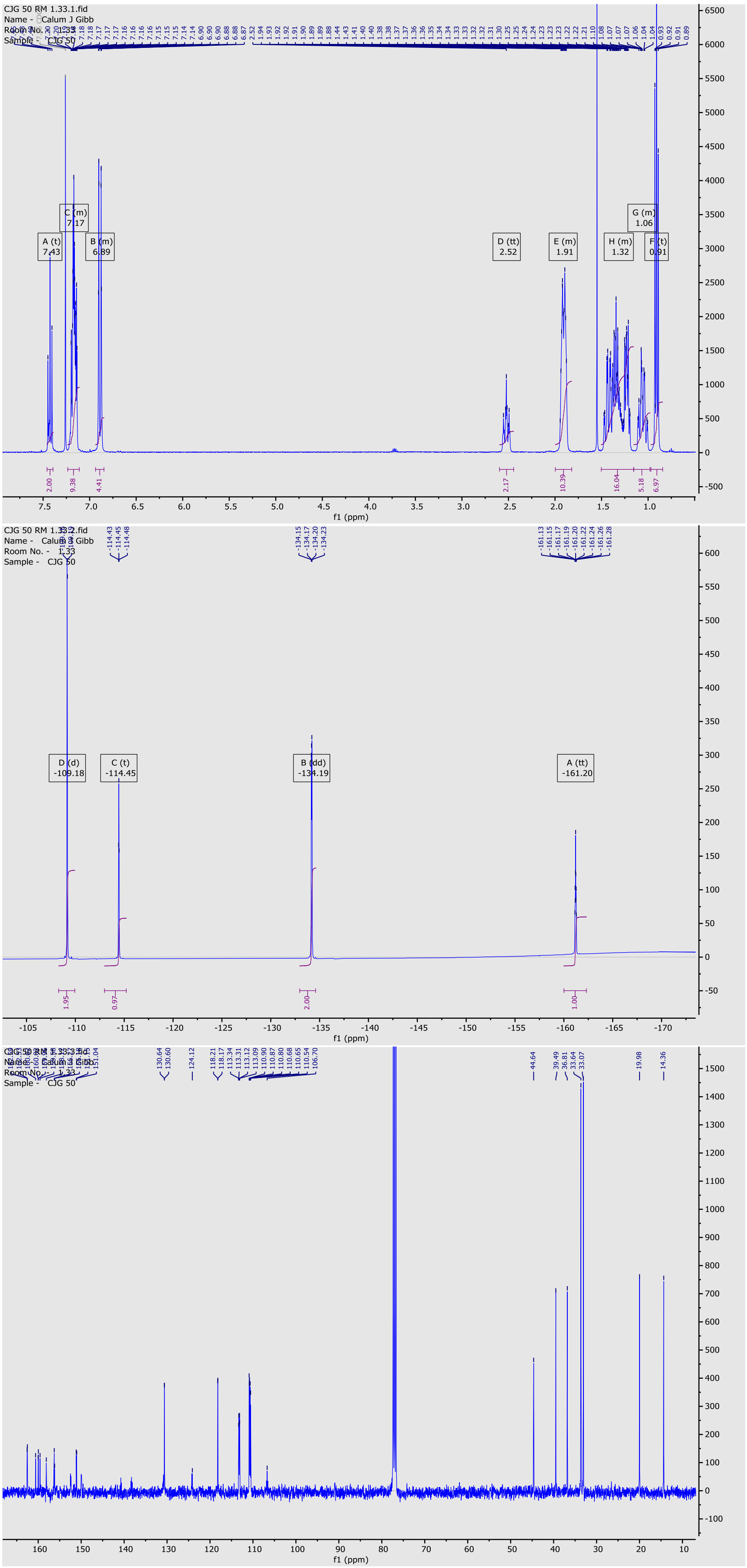

Fig. S10: E NMR spectra of **10 / ClO**: $^{1}H$ [top], $^{19}F$ [middle], and $^{13}C\{^{1}H\}$ [bottom].

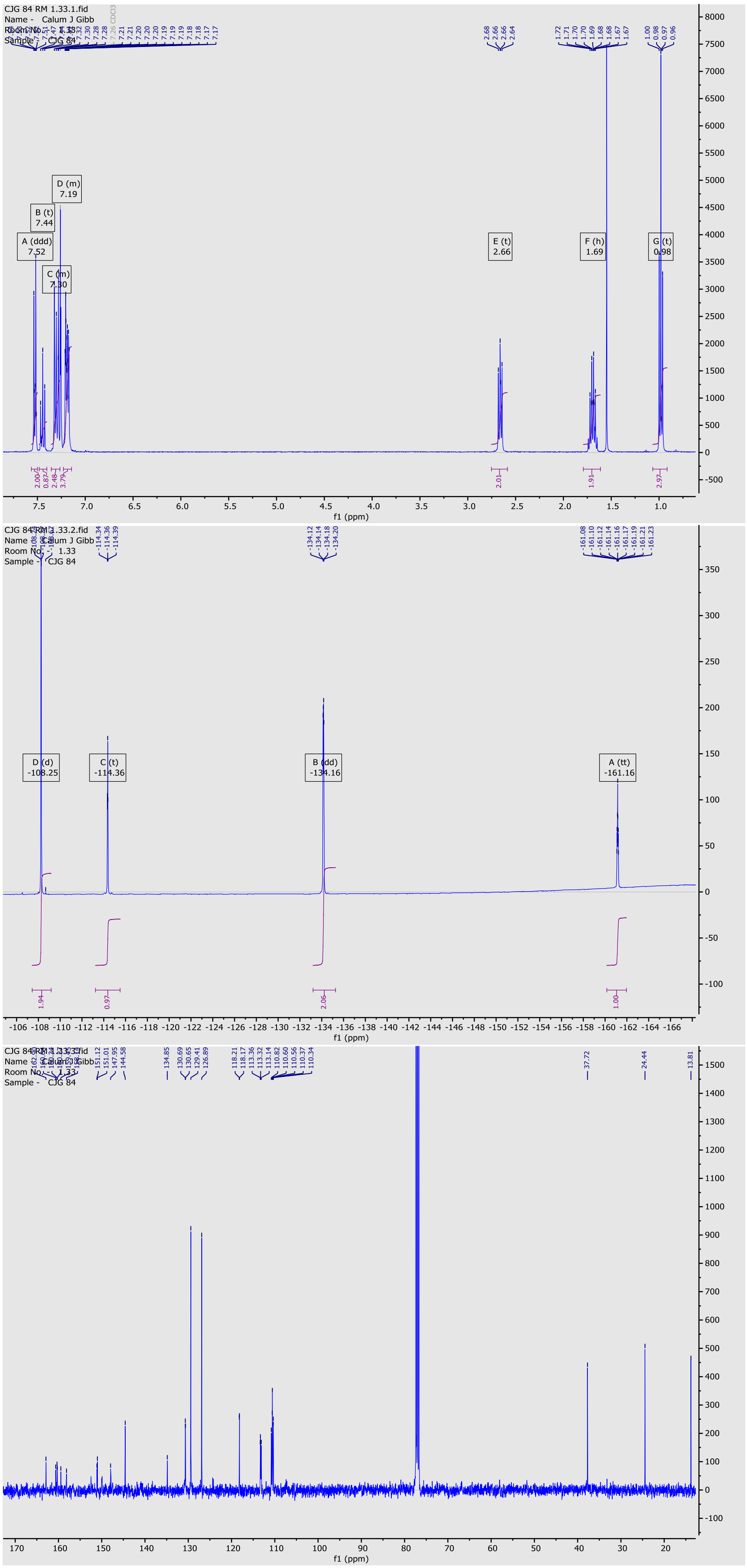

Fig. S11: E NMR spectra of **11 / PIO**: $^{1}H$ [top], $^{19}F$ [middle], and $^{13}C\{^{1}H\}$ [bottom].